\documentclass[journal,twocolumn]{IEEEtran}
\usepackage{graphicx}
\usepackage{amsmath,amsfonts,amssymb,amsthm,amstext,amscd,amsxtra,amsopn}
\usepackage{cite}
\usepackage{bm}
\usepackage{multirow}
\usepackage{esint}
\usepackage{empheq}
\usepackage{bm}
\usepackage{algorithm}
\usepackage{algpseudocode}
\usepackage{leftidx}
\usepackage{makecell}
\usepackage{stackrel}

\usepackage{subcaption}
\usepackage{xfrac}
\usepackage{hyperref}

\def\bal#1\eal{\begin{align}#1\end{align}} 
\def\suml{\sum\limits}

\def\ma{\bm{\alpha}}
\newcommand{\br}[1]{\left[#1\right]} 
\newcommand{\pr}[1]{\left(#1\right)} 
\newcommand{\cbr}[1]{\left\{#1\right\}} 
\DeclareMathOperator*{\argmin}{arg\,min} 

\def\m{\mathbf}

\def\ast{*}

\newcommand{\norm}[2]{\ensuremath{\left\|#1\right\|_{#2}}}

\newcommand {\bbmtx}{\begin{bmatrix}} 
\newcommand {\ebmtx}{\end{bmatrix}} 

\newcommand{\prox}[2]{\ensuremath{\mbox{prox}_{#1}\pr{#2}}}

\newcommand{\mat}[1]{\mathbf{#1}}

\ifCLASSINFOpdf
\else
\fi
\title{Maxwell Parallel Imaging}

\author{
    Matteo~Alessandro~Francavilla,
  Stamatios~Lefkimmiatis,
  Jorge~F.~Villena,
  ~and~Athanasios~G.~Polimeridis  
  \thanks{All authors are with Q Bio Inc., San Carlos, CA 94070, USA.}
  \thanks{Manuscript submitted to Magnetic Resonance in Medicine.}}

\begin{document}

\maketitle

\begin{abstract}

\noindent

\textit{Purpose}: To develop a general framework for Parallel Imaging (PI) with the use of Maxwell regularization for the estimation of the sensitivity maps (SMs) and constrained optimization for the parameter-free image reconstruction.

\noindent
\textit{Theory and Methods}: Certain characteristics of both the SMs and the images are routinely used to regularize the otherwise ill-posed optimization-based joint reconstruction from highly accelerated PI data. In this paper we rely on a fundamental property of SMs--they are solutions of Maxwell equations-- we construct the subspace of all possible SM distributions supported in a given field-of-view, and we promote solutions of SMs that belong in this subspace. In addition, we propose a constrained optimization scheme for the image reconstruction, as a second step, once an accurate estimation of the SMs is available. The resulting method, dubbed Maxwell Parallel Imaging (MPI), works seamlessly for arbitrary sequences (both 2D and 3D) with any trajectory and minimal calibration signals.

\noindent
\textit{Results}: The effectiveness of MPI is illustrated for a wide range of datasets with various undersampling schemes, including radial, variable-density Poisson-disc, and Cartesian, and is compared against the state-of-the-art PI methods. Finally, we include some numerical experiments that demonstrate the memory footprint reduction of the constructed Maxwell basis with the help of tensor decomposition, thus allowing the use of MPI for full 3D image reconstructions.

\noindent
\textit{Conclusions}: The MPI framework provides a physics-inspired optimization method for the accurate and efficient image reconstruction from arbitrary accelerated scans.
\end{abstract}


\section{Introduction} 
\label{sec:introduction}

Parallel Imaging (PI) is admittedly one of the most disruptive technologies in modern magnetic resonance imaging  (MRI) and probably the best example of a successful transition from academic research to widespread usage in clinic. Essentially, PI exploits the multi-physic nature of MRI and the ubiquitous use of sophisticated spatially-distributed receiving coils in order to significantly reduce the scan time. Indeed, the interplay of electrodynamics and spin-dynamics in the spatiotemporal encoding, as evinced by the bilinear form of the MR signal equation, suggests that the spatial selectivity of the receivers could be harnessed in order to reduce the time-consuming gradient encoding.

There is a plethora of PI reconstruction methods that could be roughly categorized into two main approaches: the image-space (or spatial-domain)  and the $\text{k}$-space (or spectral-domain). As main representatives of the former approach, which calls for the a-priori knowledge of the associated sensitivity maps (SMs), one can mention the pioneering works of SMASH \cite{Sodickson1997} and SENSE \cite{Pruessmann1999}. The $\text{k}$-space methods followed a few years later aiming exactly at breaking the dependence of separate pre-calibration scans, which increase the overall acquisition time and are more susceptible to motion artifacts. The beginning of those so-called auto-calibrating methods can be identified with the emergence of GRAPPA \cite{Griswold2002}, which makes use of some extra auto-calibration signals (ACS) in order to fit the kernels for approximating the missing $\text{k}$-space lines. A more detailed description of all the methods developed in the early days of PI can be found in the review paper \cite{Larkman2007}.

The first PI  techniques, both image-space and $\text{k}$-space, were geared to fast reconstruction times, allowing certain simplifications at the expense of extra pre-calibration scans or ACSs in order to transform the inherently non-linear problem into a linear one. Naturally, more sophisticated PI methods followed that consider the original bilinear form of the inverse problem at hand, incorporating the estimation of the coil SMs. The common point of the most notable among them (JSENSE \cite{Ying2007} and NLINV \cite{Uecker2008}) is the use of appropriate regularization, necessary for the otherwise ill-posed inverse problem. More specifically, in both methods the authors exploit the smoothness of the SMs by making use of a polynomial expansion for constraining the subspace of the possible solutions of the SMs in the former while applying a smoothness-enforcing regularization term in the latter. Recently, NLINV was further generalised  to a method dubbed ENLIVE with the addition of extra bilinear forms in order to account for the violation of the standard model in case of limited field-of-view (FOV)\cite{Holme2019}. Another aspect of the smoothness and the spatial selectivity of the SMs is that they also favor purely algebraic techniques based on modern numerical linear algebra algorithms that promote low-rank and subspace-specific solutions \cite{Trzasko2011, Shin2014, Haldar2014,Haldar2016}.

As the above-mentioned iterative PI reconstruction approaches started gaining more traction, the interest shifted towards the use of more expressive regularizers, ranging from ones readily available in the mathematical optimization literature \cite{Lustig2007, Huang2010, Knoll2012} to more modern data-driven variational models \cite{Knoll2020}. Again, it became clear that although the joint reconstruction of both the SMs and the images was offering certain advantages, there was a strong argument for considering the SMs estimation first and then using those SMs in the solution of the linear image reconstruction. This justifies the further proliferation of numerical methods that are tailored to the accurate estimation of the SMs \cite{Morrison2007, Jin2010, Allison2012, Uecker2014, Ma2015}; among them ESPIRiT \cite{Uecker2014} deserves a special mention as it appears to be a true workhorse and the method of choice for most of the recent studies, including the benchmark challenge for the deep-learning PI reconstruction techniques \cite{Zbontar2018}. More specifically, ESPIRiT
is based on an eigenvalue decomposition of an image-domain operator, and essentially exploits the smoothness of the SMs and the rank-deficient properties of the calibration matrix.

Evidently, the modern PI reconstruction techniques have gone a long way from the first days of accelerated MR scans and today it is quite common to use more sophisticated methods of linear and bilinear numerical optimization as well as deep-learning for reconstructing both the SMs and the images. Nevertheless, even the most effective PI methods available today are based on regularizers that are oversimplified and/or require case-dependent fine tuning of the penalty parameters. In this work, we develop a general PI framework that relies on physics-inspired regularization for the estimation of the SMs and parameter-free constrained optimization for the image reconstruction. More specifically, we note that smoothness is only one of the characteristics of SMs that depends, among other, on the scanner's main field strength. Foremost, SMs are solutions of Maxwell equations; they correspond to the magnetic fields collected by the receiving coils in the presence of the patient. Hence we choose to generate the subspace of the associated SMs (i.e. a complete numerical basis of magnetic fields in the FOV) in a patient-agnostic fashion and we proceed to the solution of the regularized bilinear optimization problem, where the SMs are expressed as arbitrary linear combinations of the elements of the Maxwell basis. In addition, we make use of a tensor compression scheme for reducing the memory footprint of the Maxwell basis in the case of 3D reconstruction. Finally, we appreciate the need for more expressive regularizers and we propose a parameter-free, constrained optimization scheme for improving the image quality when SMs are available. The effectiveness of the proposed general PI framework, dubbed Maxwell Parallel Imaging (MPI), is demonstrated for a wide range of typical sequences (both 2D and 3D) with various reduction factors (R) and ACSs.

\section{Theory and Methods}
\label{sec:theory}

\subsection{Problem Formulation}

We consider the discretized form of the PI problem, which can be described by the forward model
\bal
\m y = \m F \m S \m p + \m n,
\label{eq:forward_model}
\eal
where $\m p$ is an $ N$-dimensional vector that contains in rasterized form the samples of the unknown density to be reconstructed (e.g., a 2-D MRI slice, a 3-D MRI volume, or a 4-D MRI multi-contrast  tensor), and
$\m y, \m n \in \mathbb{C}^{K C}$ are column vectors corresponding to the $k$-space samples obtained from the $C$ receiver coils and i.i.d Gaussian noise, respectively. Furthermore, $\m S \in \mathbb{C}^{N C \times N}$ is a matrix composed as $\m S = \br{\m S_1^{\mathsf{H}} \ldots  \m S_C^{\mathsf{H}}}^{\mathsf{H}}$, where $ \m S_k \in \mathbb{C}^{N\times N}$ is a diagonal matrix constructed by the SM $\m s_k \in \mathbb{C}^N$ of the $k$th coil, $k=1, \ldots, C$, $\pr{\cdot}^\mathsf{H}$ denotes the Hermittian transpose, and $\m F \in \mathbb{C}^{K C \times N C}$ is a block diagonal matrix obtained as  $\m I_C \otimes \m F_s$, where $\m I_C \in \mathbb{R}^{C \times C}$ is the identity matrix, $\otimes$ denotes the Kronecker product, and $\m F_s \in \mathbb{C}^{K\times N}$ is the undersampled operator that provides a mapping from image space to $k$-space, with $K \le N$. The nominal R is defined as $N/K$ and corresponds to the undersampling rate of the $k$-space.

The recovery of the underlying density $\m p$ from the acquired $k$-space data $\m y$ belongs to the category of inverse problems. Due to the presence of noise $\m n$, whose exact realization is unknown, and since the operator $\m F$ is singular,  it is an ill-posed problem~\cite{Hansen2006}. This implies that in order to obtain a statistically or physically meaningful solution, we need to exploit any prior knowledge we might have about the solution. Another complicating factor that makes the recovery of $\m p$ even more challenging, is that the SMs embedded in $\m S$ are typically unknown and need to be also recovered. This results in an observation model that is not anymore linear w.r.t. the unknown quantities, but instead has the following bilinear form:
\bal
\m y = \mathcal{G}\pr{\m p, \m s_1, \ldots, \m s_C} + \m n,
\eal
where
\bal
\mathcal{G}\pr{\m x \equiv \br{\m p^\mathsf{H}, \m s_1^\mathsf{H}, \ldots, \m s_C^\mathsf{H}}^\mathsf{H}} = \bbmtx \m F_s\pr{\m s_1 \odot \m p} \\
\vdots \\
\m F_s\pr{\m s_C \odot \m p} 
\ebmtx
\eal
and $\odot$ indicates element-wise multiplication of vectors.

\subsection{Regularized Nonlinear Inversion}
\label{sec:NonLinear_recon}

One popular way to tackle the joint recovery problem of $\m p$ and $\cbr{\m s_k}_{k=1}^{C}$ is to employ the Iteratively Regularized Gauss-Newton (IRGN) method that was introduced in \cite{Bakushinsky2005} and was later used for PI reconstruction in \cite{Uecker2008}. The underlying idea of this approach consists of (a) considering the linearized approximation of the nonlinear operator $\mathcal{G}\pr{\m x}$ around some current estimate of the solution, $\mathcal{G}\pr{\m x^n + \Delta\m x} \approx \mathcal{G}\pr{\m x^n} + \m J_{\mathcal{G}}\pr{\m x^n}\Delta\m x$, where $ \m J_{\mathcal{G}}\pr{\m x^n}$ is the Jacobian of $\mathcal{G}$ evaluated at $\m x^n$, (b) minimizing an objective function of the form:
\bal
\Delta\m x^\ast = \argmin_{\Delta\m x} & \frac{1}{2}\norm{\pr{\m y - \mathcal{G}\pr{\m x^n}} - \m J_{\mathcal{G}}\pr{\m x^n}\Delta\m x}{2}^2 + \nonumber \\ 
& \mathcal{R}\pr{\m x^n + \Delta\m x},
\label{eq:IRGN}
\eal
where $\mathcal{R}\pr{\cdot}$ is a regularization functional and (c) updating the current estimate as $\m x^{n+1} = \m x^{n} + \gamma \Delta\m x^{\ast}$, where $\gamma$ can be computed using a line-search strategy.  

Initially, in ~\cite{Uecker2008} the authors considered using the regularizer $\mathcal{R}\pr{\m x^n} = \alpha_n\norm{\m p^n}{2}^2 + \beta_n\suml_{k=1}^C\norm{\m W \tilde{\m F}\m s_k^n}{2}^2$, where $\tilde{\m F}\in\mathbb{C}^{N\times N}$ is the DFT matrix, $\m W\in \mathbb{R}^{N\times N}$ is a weighting diagonal matrix and $\alpha_n, \beta_n \ge 0$. Note that, given that the SMs are expected to be smooth, the second term of this regularizer penalizes their high-frequency content. Later, in ~\cite{Knoll2012} the authors replaced the Tikhonov regularizer on the density, $\norm{\m p^n}{2}^2$, with non-quadratic regularizers that can better model certain properties of the underlying density, at the cost of a more involved minimization strategy; the minimizer of Eq.~\eqref{eq:IRGN} cannot be derived anymore as the solution of a system of linear equations and more advanced convex optimization techniques must be employed.

In this work, we also rely on the IRGN method as an initial step that provides an estimate of the unknown SMs. Then in a second step, as we describe later, we use the estimated SMs in order to recover a high-quality estimate of the underlying density by solving the linear inverse problem of Eq.~\eqref{eq:forward_model}. We note that such a two-step strategy has been regularly followed in other image processing applications, such as blind deconvolution~\cite{Levin2011}, where apart from the underlying image the degradation operator is also unknown. Unlike Refs. \cite{Uecker2008, Knoll2012}, we consider a modified version of the objective function in~\eqref{eq:IRGN}, where instead of regularizing directly the $k$th SM $\m s_k$, we penalize its expansion coefficients $\ma_k$ on a predefined subset of basis vectors $\m U\in \mathbb{C}^{N\times q}$. Details on the construction of an appropriate physics-inspired basis are provided in the following Sections. In particular, we express each SM as $\m s_k \approx \m U\ma_k$ and thus, our observation model takes the form:
\bal
\m y = \mathcal{G}\pr{\m p, \ma_1, \ldots, \ma_C} + \m n, 
\label{eq:forward_model_maxwell}
\eal
where $\mathcal{G}\pr{\hat{\m x} \equiv \br{\m p^\mathsf{H}, \ma_1^\mathsf{H}, \ldots, \ma_C^\mathsf{H}}^\mathsf{H}} = \bbmtx \m F_s\pr{\m U\ma_1 \odot \m p} \vspace{-.2cm}\\
\vdots \vspace{-.2cm}\\
\m F_s\pr{\m U\ma_C \odot \m p}
\ebmtx.$

Then, we seek for the solution of the following minimization problem:
\bal
\Delta\hat{\m x}^\ast = \argmin_{\Delta\m p, \Delta\ma_1, \ldots, \Delta\ma_C} & \frac{1}{2}\norm{\pr{\m y - \mathcal{G}\pr{\hat{\m x}^n}} - \m J_{\mathcal{G}}\pr{\hat{\m x}^n}\Delta\hat{\m x}}{2}^2 + \nonumber \\
& \alpha_n\norm{\m p^n + \Delta\m p}{2}^2 + \nonumber \\
& \beta_n\suml_{k=1}^C\norm{\ma_k^n +\Delta \ma_k}{2}^2,
\label{eq:IRGN_modified}
\eal
where we impose an $\ell_2$-squared penalty both on the density and the expansion coefficients of the SMs. It is worth noticing that by estimating the expansion coefficients $\ma$ instead of the SMs themselves and since the coils are represented in a \emph{reduced order model}, i.e. $N >\!\!> q$, the solution of Eq.~\eqref{eq:IRGN_modified} corresponds to that of an over-determined problem. Due to the quadratic form of the objective function to be minimized in  Eq.~\eqref{eq:IRGN_modified}, the solution can be derived by solving the relevant normal equations using the conjugate gradient method~\cite{Shewchuk1994}. This requires the ability to compute the matrix-vector products of the Jacobian of $\mathcal{G}$ and its adjoint, with a vector. These products are computed as follows: 
\bal
\m J_{\mathcal{G}}\pr{\hat{\m x}}\br{\begin{smallmatrix} \Delta\m p \\ \Delta\ma_1 \\ \vdots \\ \Delta\ma_C\end{smallmatrix}} & = &
 \mathcal{G}\pr{\br{\Delta\m p^\mathsf{H}, \ma_1^\mathsf{H}, \ldots, \ma_C^\mathsf{H}}^\mathsf{H}
} + \notag \\
& & \mathcal{G}\pr{\br{\m p^\mathsf{H}, \Delta \ma_1^\mathsf{H}, \ldots, \Delta \ma_C^\mathsf{H}}^\mathsf{H}} \notag \\
& = & \br{\begin{matrix}\m F_s\pr{\m U\ma_1 \odot \Delta\m p + \m p \odot \m U\Delta\ma_1} \vspace{-.2cm}\\
\vdots \vspace{-.2cm}\\
\m F_s\pr{\m U\ma_C \odot \Delta\m p + \m p \odot \m U\Delta\ma_C}
\end{matrix}}
\eal
and
\bal
\m J_{\mathcal{G}}^\mathsf{H}\pr{\hat{\m x}}\br{\begin{matrix} \vspace{-.1cm}\m y_1 \\ \vdots \vspace{-.2cm}\\ \m y_C\end{matrix}} =
\br{\begin{matrix}
\suml_{k=1}^C \pr{\m U\ma_k}^\mathsf{H} \odot \m F_s^\mathsf{H}\m y_k \\
\m U^\mathsf{H}\pr{\m p^\mathsf{H}\odot \m F_s^\mathsf{H}\m y_1} \vspace{-.2cm}\\
\vdots \vspace{-.2cm}\\
\m U^\mathsf{H}\pr{\m p^\mathsf{H}\odot \m F_s^\mathsf{H}\m y_C}
\end{matrix}}.
\notag
\eal
Finally, it's worth noting that the same procedure can be followed in the case of dynamic or multi-contrast PI data, expecting that the associated artifacts will be captured by the density while the estimation of the (constrained) SMs will remain unaffected, especially when using some average of the $k$-space data.

\subsection{Regularized Density Reconstruction via Constrained Optimization}
\label{sec:SENSE}

While the regularization applied on $\m p$ in Eq.~\eqref{eq:IRGN_modified} is rather plain and thus, not capable of modeling complex properties of the underlying density, it allows us to perform a joint reconstruction of the sensitivities and the density without having to rely upon a computationally heavy and time consuming minimization scheme. Furthermore, due to the implicit regularization of the SMs, by expressing them in terms of a proper basis expansion, and the explicit Tikhonov regularization of the corresponding expansion coefficients, we expect that most of the reconstruction errors will be accumulated in the recovered density, while the unknown SMs will be more accurately restored. 

Having this in mind, we use the estimated SMs, discard the estimated density and solve the linear inverse problem described in Eq.~\eqref{eq:forward_model}. Hence, we obtain the final density estimate as the minimizer of the following constrained optimization problem:
\bal
\m p^\ast = \argmin_{\substack{\m p\in \mathbb{C}^N \\ \norm{\m y_k - \m F_s\m S_k\m p}{2}\le \varepsilon_k, \forall k}}\mathcal{R}\pr{\m p},
\label{eq:constrained_min}
\eal
where $\varepsilon_k$ is a scalar that is proportional to the standard deviation of the complex Gaussian noise realization that degrades the $k$-space measurements acquired from the $k$th coil. While for the experiments that we report in this work, we have considered Total Variation~\cite{Rudin1992} as the regularization functional $\mathcal{R}\pr{\m p}$ of choice, the minimization strategy that we outline next can be also used without modifications when different and more expressive regularizers are considered, such as the Structure Tensor Total Variation (STV) ~\cite{Lefkimmiatis2015Ja} and it's non-local extension~\cite{Lefkimmiatis2015Jb} or the Hessian-Schatten norm regularizers of~\cite{Lefkimmiatis2013J}. We also note that one particular advantage of the above constrained problem formulation, compared to the unconstrained minimization approach that is typically pursued in PI reconstruction, is that in this case there is no need of fine-tuning any regularization penalty parameter, which in practice is not a straightforward task and requires a certain level of experience from the user. The only parameters involved in the above formulation, are the scalars $\varepsilon_k$ which can be directly estimated from the $k$-space measurements.

Now, let us first note that the constrained formulation of Eq.~\eqref{eq:constrained_min} can be equivalently expressed in the unconstrained form 
\bal
\m p^\ast = \argmin_{\m p\in \mathbb{C}^N}\mathcal{R}\pr{\m p} + \suml_{k=1}^C\iota_{\mathcal{C}\pr{\m y_k, \varepsilon_k}}\pr{\m F_s\m S_k \m p}, 
\label{eq:unconstrained_min}
\eal
where 
\bal
\iota_{\mathcal{C}\pr{\m y_k, \varepsilon_k}}\pr{\m z} = \begin{cases} 0,& \mbox{if } \norm{\m y_k - \m z}{2}\le \varepsilon_k \\ \infty, &\mbox{otherwise} \end{cases}
\notag
\eal
is an indicator function which ensures that the imposed constraints are satisfied by the solution.  Next, since the transformed problem is still hard to solve directly, we rely on the Alternating Direction Method of Multipliers (ADMM)~\cite{Esser2009, Boyd2011}. The strategy of ADMM is to split the original problem in smaller and easier ones to solve, by decoupling the different terms of the objective function. Based on this idea and following a similar splitting strategy as in~\cite{Lefkimmiatis2013Jb}, we instead consider the problem
\bal
\min_{\m A\m p + \m B\m z = \m 0}\mathcal{R}\pr{\m z_0} + \suml_{k=1}^C\iota_{\mathcal{C}\pr{\m y_k, \varepsilon_k}}\pr{\m z_k}, 
\eal
where $\m A =  \br{\m I_N, \pr{\m F_s \m S_1}^\mathsf{H}, \ldots,  \pr{\m F_s \m S_C}^{\mathsf{H}}}^{\mathsf{H}}$, $\m B = - \m I_{\pr{N+K C}}$ and $\m z = \br{\m z_0^\mathsf{H} \ldots \m z_C^\mathsf{H}}^\mathsf{H} \in \mathbb{C}^{N+K C}$. Then, using the \emph{scaled form} of ADMM~\cite{Boyd2011} we obtain the solution to our original problem of Eq.~\eqref{eq:unconstrained_min} in an iterative way, where each iteration involves the following update steps:
\bal
\m z_0^{n+1} &= \prox{1/\rho\cdot\mathcal{R}}{\m z_0^n - \pr{\m p^n + \m u_0^n}}, \notag\\
\m z_k^{n+1} & = \bm{\Pi}_{\mathcal{C}\pr{\m y_k, \varepsilon_k}}\pr{\m F_s\m S_k\m p^n + \m u_k^n}, \forall k = 1, \ldots, C, \notag\\
\m p^{n+1} & = \pr{\m I_N + \suml_{k=1}^C \m S_k^\mathsf{H}\m F_s^\mathsf{H}\m F_s \m S_k}^{-1} \notag\\ 
& \pr{\m z_0^{n+1} -\m u_0^{n} + \suml_{k=1}^C \m S_k^\mathsf{H}\m F_s^\mathsf{H}\pr{\m z_k^{n+1}-\m u_k^{n}}},\notag\\
\m u^{n+1} &= \m u^n + \m A \m p - \m z.
\label{eq:algo}
\eal
In Eq.~\eqref{eq:algo} we have that $\bm{\Pi}_{\mathcal{C}\pr{\m y, \varepsilon}}\pr{\m z} = \m y + \frac{\varepsilon\pr{\m z - \m y}}{\max\pr{\norm{\m z-\m y}{2}, \varepsilon}}$, $\prox{1/\rho\mathcal{R}}{\m z} = \argmin_{\m x} \rho/2 \norm{\m x - \m z}{2}^2 + \mathcal{R}\pr{\m x}$ is the proximal operator~\cite{Combettes2005} of the regularizer $\mathcal{R}\pr{\cdot}/\rho$, $\m u = \br{\m u_0^\mathsf{H} \ldots \m u_C^\mathsf{H}}^\mathsf{H} \in \mathbb{C}^{N+K C}$ are the \emph{dual variables},  and $\rho$ is the ADMM penalty parameter. In order to avoid fine-tuning the ADMM penalty parameter $\rho$, whose value can affect the convergence rate of the minimization algorithm, we adaptively choose its value in each iteration so as to balance the primal and dual residuals (see~\cite{Boyd2011} for their definitions), as proposed in~\cite{He2000}.

The linear reconstruction algorithm that we proposed above is general enough to accommodate for different MRI acquisition modalities. In particular, the steps described in Eq.~\eqref{eq:algo} can also be applied when multicontrast or dynamic MRI are considered. The main difference is that for multicontrast MRI, the regularizer $\mathcal{R}\pr{\cdot}$ instead of being applied only on the spatial dimensions of the underlying density, it should also act on the different contrast channels so that it accounts for the dependencies that exist among them. A possible regularizer that can be used for this task is the Vectorial Total Variation~\cite{Blomgren1998}. As far as it concerns the dynamic MRI case, the solution can be expressed as the minimization of a constrained problem very similar to the one in Eq.~\eqref{eq:constrained_min},
\bal
\m p^\ast = \argmin_{\substack{\m p\in \mathbb{C}^{N\times T} \\ \norm{\m y_{k,t} - \m F_s^t\m S_{k}\m p_t}{2}\le \varepsilon_{k,t}, \forall k, t}}\mathcal{R}\pr{\m p},
\label{eq:constrained_min_dynamic}
\eal
where $\m p = \br{\m p_1 \ldots \m p_T}$, $\m y_{k,t}$, $\varepsilon_{k,t}$ correspond to the $k$-space measurements and a scalar proportional to the standard deviation of the noise from the $k$th coil and the $t$th time instance, respectively, $\m F_k^t$ is the undersampled mapping operator used at time instance $t=1,\ldots, T$ and $\mathcal{R}\pr{\m p}$ is a spatiotemporal regularizer. Then, one can follow the strategy described above to obtain a slightly modified version of the algorithmic steps provided in Eq.~\eqref{eq:algo}.

\subsection{Maxwell Regularization}
\label{sec:basis}

A key ingredient of the proposed non linear inversion scheme is the physics inspired regularization of the coil model. Because SMs $\mat{s}_{k}$ are solutions to Maxwell equations, we propose to constrain the solution space of the imaging problem to a subspace where SMs are indeed solutions to Maxwell equations, and conjecture that it is possible to express each SM as
$
\mat{s}_{k} \approx \mat{U}_h \alpha_{k}
$, 
where $\alpha_{k} \in \mathbb{C}^{q}$ is a column vector collecting the expansion coefficients of the $k$th coil, and $\mat{U}_h \in \mathbb{C}^{N \times q}$ is a proper change of basis matrix, referred to as Maxwell basis in the following. The dimension $q$ of the basis will play an important role in controlling the accuracy of the representation and the regularization properties.

Because the basis $\mat{U}_h$ collects solutions of Maxwell equations, a scheme for solving Maxwell equations within the FOV is a prerequisite.
One approach is based on Love's form of the field equivalence theorem \cite{Love1901, Ishimaru1978}:
fields inside a source-free volume are fully determined if the tangential electromagnetic (EM) fields on the boundary of the volume are known.
Following this idea, the problem of finding volumetric fields inside the source-free FOV is conveniently addressed as a two-step procedure: first, solve for equivalent electric ($\mat{j}$) and magnetic ($\mat{m}$) currents on the boundary. Subsequently, EM fields $\mat{e}$ and $\mat{h}$ inside the FOV are expanded as
\begin{equation}
\begin{bmatrix}
\mat{e} \\[0.5em]
\mat{h}
\end{bmatrix}
=
\begin{bmatrix}
\mathcal{K}^{ej} & \mathcal{K}^{em} \\[0.5em]
\mathcal{K}^{hj} & \mathcal{K}^{hm}
\end{bmatrix}
\begin{bmatrix}
\mat{j} \\[0.5em]
\mat{m}
\end{bmatrix},
\label{eq:inc_operators}
\end{equation}
where $\mathcal{K}^{\alpha\beta}$ is a linear integro-differential operator defined as
\begin{equation}
    \mathcal{K}^{\alpha\beta} \, \mat{f} = \int_{\mathbb{R}^3} \mathcal{G}^{\alpha\beta} \left( \mat{r}, \mat{r}^\prime \right) \cdot \mat{f} \left( \mat{r}^\prime \right) d\mat{r}^\prime,
\end{equation}
and $\mathcal{G}^{\alpha\beta}$ is the dyadic Green function, mapping $\beta$-kind currents to $\alpha$-kind fields. 

Note that $\mat{j}$ and $\mat{m}$ are only proxies for computing $\mat{h}$: we are interested in finding a basis to represent a basis for all possible realizations of $\mat{h}$. This reflects into the need of spanning the range of the integral operators $\mathcal{K}^{hj}$ and $\mathcal{K}^{hm}$, and not in a particular solution of Eq.~\eqref{eq:inc_operators}.

\begin{figure*}[!!h]
\centering
    {
   \includegraphics[width=\textwidth]{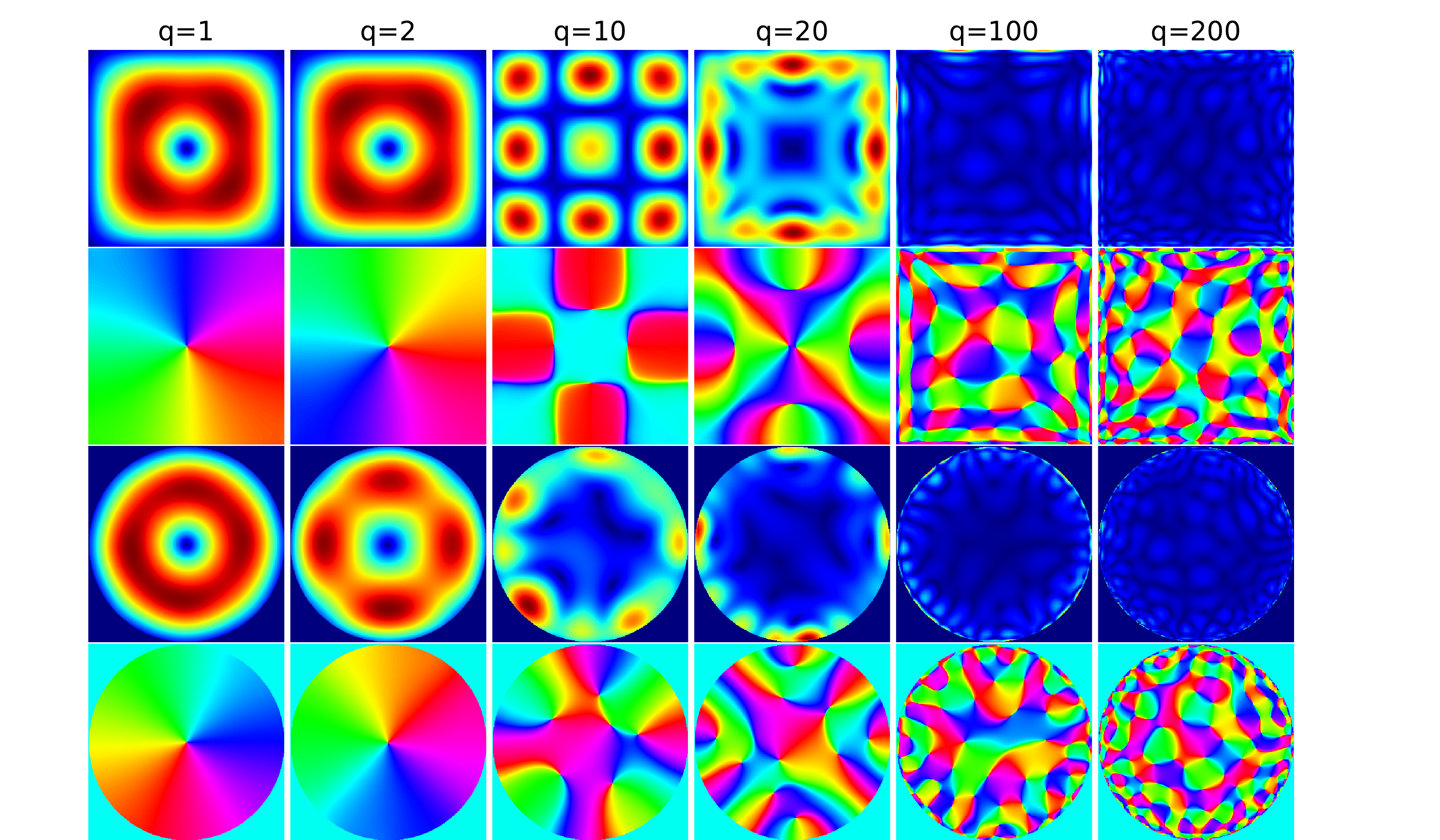}
    }
\caption{Elements of the basis: pictorial representation of a selection of elements of the magnetic field basis $\mat{U}_h$ adopted to represent coil SMs.
The Maxwell basis $\mat{U}_h$ is computed for a 2D FOV and matrix of size 220x220mm and 320x320, respectively. The number of random excitations to sample the range of the integral operator via randomized SVD is 500. The basis is computed over a support covering the full FOV (top two rows), and with a circular support with a diameter of 220mm (bottom two rows). For the two scenarios, the magnitude (rows 1 and 3, arbitrary units) and phase (rows 2 and 4, represented on a cyclical HSV color scale to suppress $2\pi$ phase jumps) are plotted for the basis vectors $\mat{u}_q$, with $q \in \left[ 1, 2, 10, 20, 100, 200 \right]$. Because of the SVD-based scheme, $\mat{U}_h$ is a spectral basis: vectors are ordered according to their spatial frequency content, higher index vectors modelling faster variations (from left to right). As a consequence, increasing the dimension of the basis increases the high-frequency components of the SM that can be captured by $\mat{U}_h$.}
\end{figure*}

One way to obtain an orthonormal basis $\mat{U}_h$ is to compute the left singular vectors of 
\begin{equation}
\label{eq:em_discretized}
    \mat{K} = 
        \begin{bmatrix}
            \mat{K}^{hj}           &   \mat{K}^{hm}
        \end{bmatrix},
\end{equation}
where $\mat{K}^{hj}$ and $\mat{K}^{hm}$ are the discrete representations of operators $\mathcal{K}^{hj}$ and $\mathcal{K}^{hm}$, respectively.
Then, if $\mat{K} = \mat{U} \mat{\Sigma} \mat{V}^\mathsf{H}$ is a Singular Value Decomposition (SVD) of $\mat{K}$, $\mat{U}$ is an orthonormal basis for the range of $\mat{K}$.
A significant advantage of obtaining the basis via SVD is that it also provides the optimal low-rank approximation of $\mat{K}$. If $\mat{K} \in \mathbb{C}^{m \times n}$, among all matrices $\mat{B} \in \mathbb{C}^{m \times n}$ with rank $k$, the one obtained by truncated SVD is the one with minimum error w.r.t. the spectral norm,
\begin{equation}
    \left\lVert \mat{K} - \mat{U}_k \mat{\Sigma}_k \mat{V}_k^\mathsf{H} \right\rVert_{2} = \sigma_{k+1} \left\lVert \mat{K} \right\rVert_{2},
\end{equation}
where only the $k$ column vectors of $\mat{U}$ and $\mat{V}$ corresponding to the $k$ largest singular values are kept, and $\sigma_{k+1}$ is the $( \! k \! + \! 1 \!)$-th singular value of $\mat{K}$. In other terms, $\mat{K}_k \equiv \mat{U}_k \mat{\Sigma}_k \mat{V}_k^\mathsf{H}$ approximates $\mat{K}$ with error $\sigma_{k+1}$.
In turn, by defining $\mat{U}_h \equiv \mat{U}_k$ we have an orthonormal basis to approximate the range of $\mat{K}$ with error $\sigma_{k+1}$.

Unfortunately, evaluation of the dyadic Green functions in Eq.~\eqref{eq:inc_operators} requires knowledge of the object to be imaged: this implies that $\mat{U}_h$ is acquisition dependent, which would clearly be a major limitation. 
However, in view of the investigations documented in \cite{Vaidya2016} and references therein, at MRI frequencies the magnetic field is only slightly perturbed by the biological tissue, due to its weakly magnetic properties and the relatively small (in terms of electric length) FOV.  We then conjecture that, in the absence of fast spatial variations in the magnetic field, the problem can be simplified by a homogeneous medium problem, and one can rewrite the field equation for $\mat{h}$ in Eq.~\eqref{eq:inc_operators} in terms of the free-space scalar Green function $g \left( \mat{r}, \mat{r}^\prime \right) = \dfrac{e^{-jk \left| \mat{r} - \mat{r}^\prime \right| }}{4 \pi \left| \mat{r} - \mat{r}^\prime \right|}$, with $k = \omega \sqrt{\epsilon_0 \mu_0}$ the wavenumber in vacuum:
\begin{eqnarray}
\mathcal{G}^{hj} \left( \mat{r}, \mat{r}^\prime \right) = \nabla g \left( \mat{r}, \mat{r}^\prime \right) \times \mathcal{I}, \\
\mathcal{G}^{hm} \left( \mat{r}, \mat{r}^\prime \right) = \frac{1}{j \omega \mu} \left( \nabla\nabla + k^2 \right) g \left( \mat{r}, \mat{r}^\prime \right).
\end{eqnarray}
This is crucial for the practical applicability of the method: because the basis is computed in the absence of the biological tissue, it is universally applicable to all imaging problems sharing the same FOV. In other words, the basis is pre-computed offline for a few FOVs of interest, given only the dimensions of the FOV and the target resolution. In practice, this is achieved via a numerical discretization of Eq.~\eqref{eq:inc_operators}. More specifically, in this work we obtain the discretized linear operator in matrix form with the help of  the open-source package MARIE \cite{MARIE}, based on the methods presented in \cite{Polimeridis2014, Villena2016}.

Finally, we observe that computing the SVD of $\mat{K}$ is not feasible for practical problems, due to the extremely large size of $\mat{K}$. As a matter of fact, $\mat{K}$ is only known via its sparse factorization. A remedy to this is to resort to the so-called randomized matrix decompositions \cite{Liberty20167, Halko2011}, numerical techniques that have attracted growing interest recently thanks to their effectiveness in computing low-rank approximations of very large matrices. Because the range of a linear operator can be sampled with arbitrary precision if the images of independent and random source distributions are known, by exciting dipoles located on the boundary with random amplitudes and phases it is possible to sample the left subspace of $\mat{K}$ without actually building it. Finally, because the detected MRI signal is a circular polarization of the magnetic field $\mat{h}$, the subspace is further restricted to span only circularly polarized fields.

Figure~1 exemplifies the elements of a typical Maxwell basis over square and circular supports. The randomized SVD based approach guarantees that the basis vectors possess a spatial frequency content growing with the index of the basis vector: increasing the dimension of the basis increases the high-frequency components of the SM that can be captured by $\mat{U}_h$. Consequently, the low-pass filtering properties behave as a regularizer for the inverse problem. The representation properties of the basis are demonstrated in Figure~2, where the capability to expand known synthetic 2D SMs via the basis is analyzed. The convergence of the error of the expansion of a known SM is shown as a function of the basis dimension $q$, proving that by increasing the dimension of the basis it is possible to control the accuracy of the representation.

\begin{figure}[!htbp]
\centering
    {
   \includegraphics[width=\linewidth]{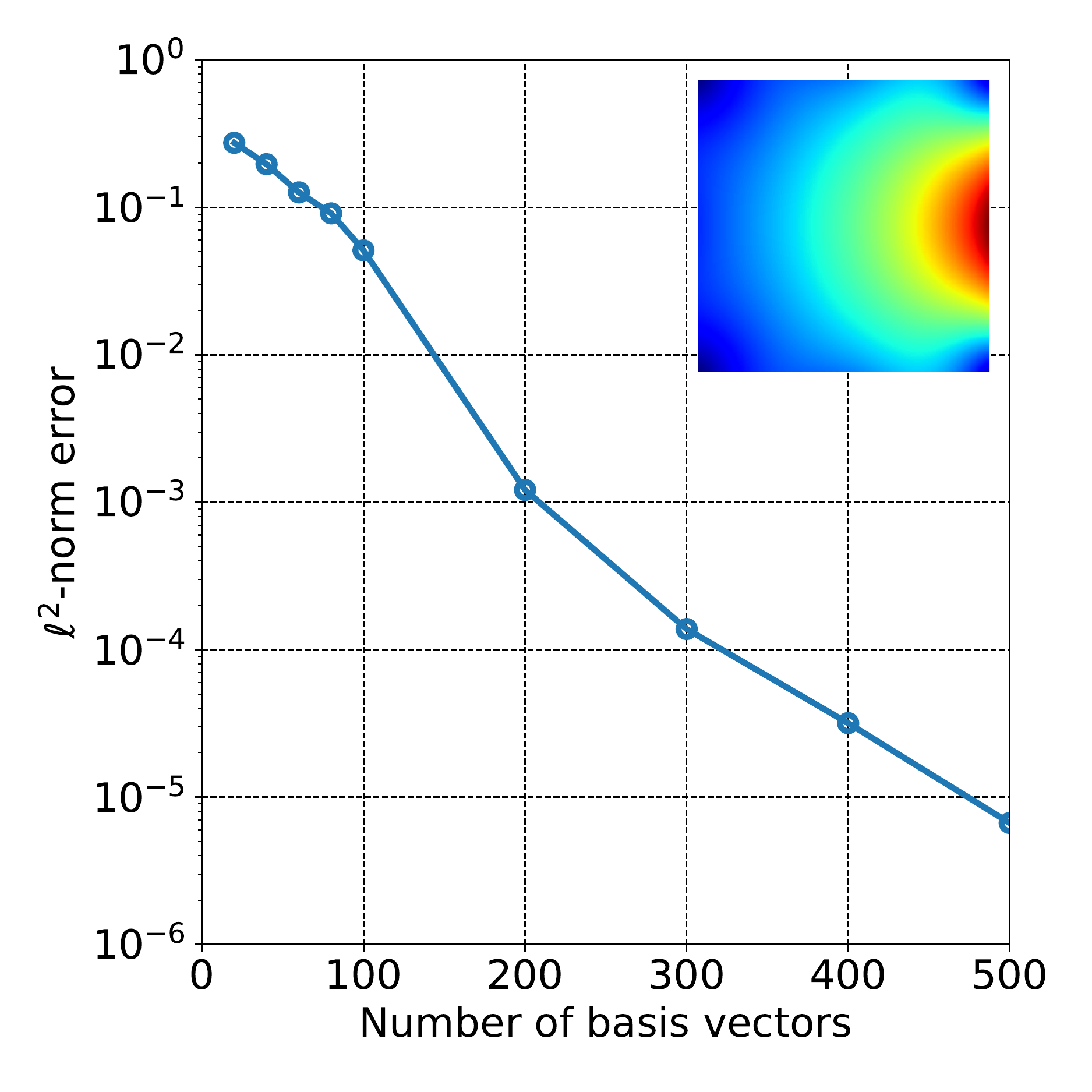}
    }
\caption{A set of 8 synthetic SMs, each of size 256x256 pixels, is generated via an open-source Python package \cite{sigpy}. The magnetic field basis is then used to represent the same set of SMs, with dimension of the basis (number of basis vectors) ranging from 20 to 500. If $\mat{U}_h$ is the matrix storing the basis, and $\mat{s}_i$ is the SM of the $i$th coil unrolled as a column vector, we denote the projection onto the basis as $\widetilde{\mat{s}}_i \approx \mat{U}_h \left( \mat{U}_h^H \mat{U}_h \right)^{-1} \mat{U}_h^{H} \mat{s}_i $.
For each coil $i$ and each size $q$ of the basis, the projection error is computed as $ e_{q,i} = \frac{\left\lVert \widetilde{\mat{c}}_i - \mat{c}_i \right\rVert_{2}}{\left\lVert \mat{c}_i \right\rVert_{2}} $, and the largest error among all coils $\displaystyle \max_{i} \left( e_{q,i} \right) $ is displayed as a function of $q$.
Inset: the SM magnitude of one of the coils is shown.}
\end{figure}

\subsection{A Compression Scheme for Maxwell Basis} 
\label{sec:tucker}

The proposed method is valid for fully 3D problems, i.e. for 3D acquisitions over volumetric FOVs, or can be restricted to 2D problems. In the latter case, the range of $\mat{K}$, and thus the support of the basis, is restricted to a single slice. 
On the other hand, when the problem is fully 3D, storage requirements for the basis itself can be a limitation. 
Because the adopted discretization is a finite-element basis, each entry of one basis vector is proportional to the field intensity sampled at the centroid of a voxel: each column of $\mat{U}_h$ can be reshaped as a three-dimensional tensor representing a three-dimensional field distribution $\mathcal{U}_i = r \left( u_i \right)$, with
\begin{equation*}
    r: \mathbb{C}^N \longrightarrow \mathbb{C}^{n_1 \times n_2 \times n_3}, \quad N = n_1 n_2 n_3
\end{equation*}
being a reshape operator reordering entries of a column vector onto a Cartesian grid.

Here we follow the idea pioneered by Tucker in \cite{Tucker1966}, which introduces a high-order singular value decomposition known as Tucker decomposition. More specifically, Tucker decomposition is used to decompose a tensor $\mathcal{T} \in \mathbb{C}^{n_1 \times n_2 \ldots \times n_M }$ to a core tensor $\mathcal{G} \in \mathbb{C}^{n_1 \times n_2 \ldots \times n_M}$ multiplied by a unitary matrix $\mat{U}_k \in \mathbb{C}^{n_k \times n_k}$ along each mode $k$. In three dimensions:
\begin{equation} \label{eq:tucker_decomposition_3d}
    \mathcal{T} = \mathcal{G} \times_1 \mat{U}_1 \times_2 \mat{U}_2 \times_3 \mat{U}_3.
\end{equation}
$\mathcal{A} \times_k \mat{B}$ denotes the $k$-mode product between a tensor $\mathcal{A}$ and a matrix $\mat{B}$ obtained as a convolution along the $k$th axis. For instance, the 1-mode product is defined as:
\begin{equation*}
    \mathcal{C} = \mathcal{A} \times_1 \mat{B}, \qquad \mathcal{C}_{ijk} = \sum_{p=1}^{n_1} \mathcal{A}_{pjk} \mat{B}_{ip}
\end{equation*}
For an intuition of the decomposition, we find it useful to pictorially visualize the 3D version as in Figure~3. If we accept an approximation of $\mathcal{T}$, the size of $\mathcal{G}$ (the multilinear ranks of the decomposition) can be much smaller than the size of $\mathcal{T}$, hence the compression. Similarly to the truncated SVD, we can truncate the expansion in Eq.~\eqref{eq:tucker_decomposition_3d} with a reduced core tensor $\mathcal{\widetilde{G}} \in \mathbb{C}^{r_1 \times r_2 \times r_3}$ and reduced unitary matrices $\mat{\widetilde{U}}_k \in \mathbb{C}^{r_k \times n_k}$, with $r_k \le n_k$:
\begin{equation} \label{eq:truncated_tucker}
    \mathcal{T} \approx \mathcal{\widetilde{G}} \times_1 \mat{\widetilde{U}}_1 \times_2 \mat{\widetilde{U}}_2 \times_3 \mat{\widetilde{U}}_3.
\end{equation}
A key feature of the expansion in Eq.~\eqref{eq:truncated_tucker} is that it can be obtained with controlled accuracy, i.e. given $\varepsilon > 0$ it is possible to find a Tucker expansion of $\mathcal{T}$ such that
\begin{equation} \label{eq:tucker_error}
    \left\lVert \mathcal{T} - \mathcal{\widetilde{G}} \times_1 \mat{\widetilde{U}}_1 \times_2 \mat{\widetilde{U}}_2 \times_3 \mat{\widetilde{U}}_3 \right\rVert_{2} < \varepsilon \left\lVert \mathcal{T} \right\rVert_{2}
\end{equation}
For an overview of the algorithms to obtain a compressed Tucker representation, the interested reader is referred to \cite{Tucker1966, Giannakopoulos2019, rabanser2017introduction} and references therein.

\begin{figure}[!htbp]
\centering
    {
   \includegraphics[width=\linewidth]{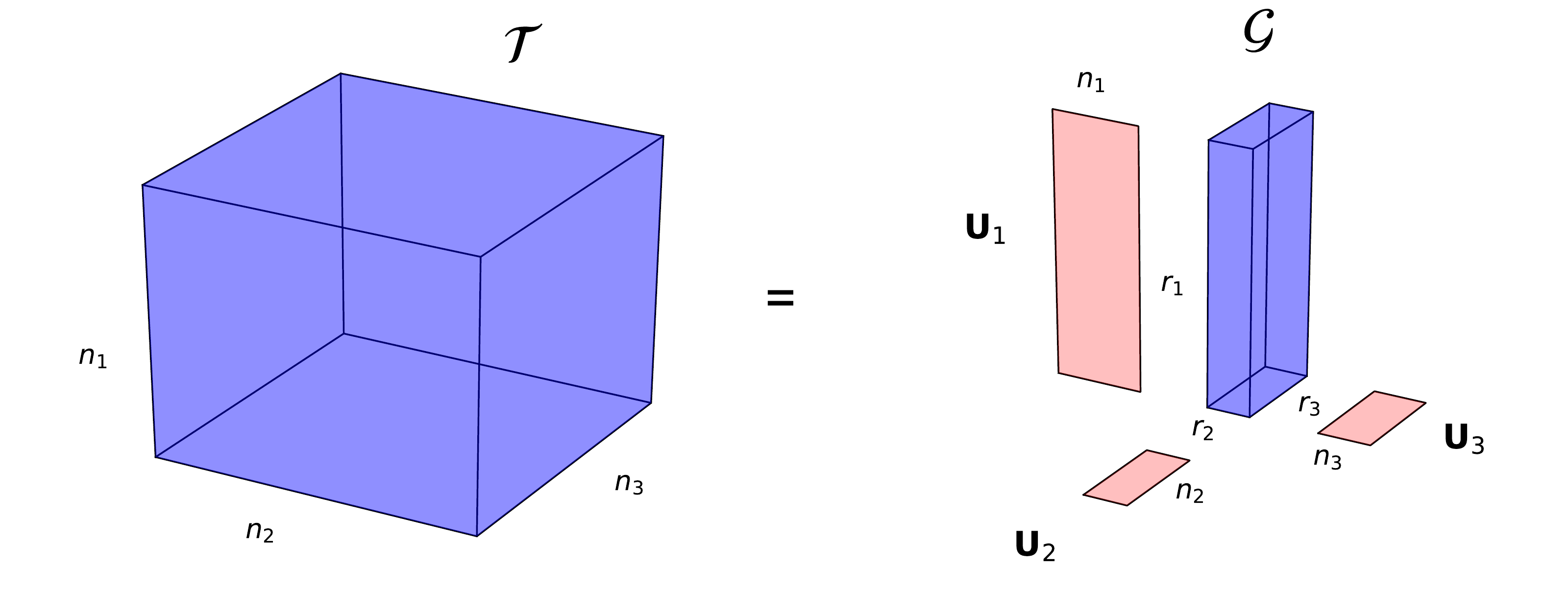}
    }
\caption{Pictorial representation of the Tucker decomposition of a 3D tensor $\mathcal{T} = \mathcal{G} \times_1 \mat{U}_1 \times_2 \mat{U}_2 \times_3 \mat{U}_3$. For a better understanding of the $k-$mode product $\times_k$, it is useful to realize that it amounts to multiplying each mode-$k$ fiber of the core tensor $\mathcal{G}$ by the matrix $\mat{U}_k$, i.e. it is a convolution along the $k$th axis of $\mathcal{G}$. If the tensor $\mathcal{T}$ is rank deficient, we have that $r_k < n_k$ and the decomposition results in a compression. When $\mathcal{T}$ is full rank we can introduce an approximated decomposition of $\mathcal{T}$  up to an arbitrary accuracy $\varepsilon$, by truncating the ranks $r_k$ such that $\left\lVert \mathcal{T} - \mathcal{\widetilde{G}} \times_1 \mat{\widetilde{U}}_1 \times_2 \mat{\widetilde{U}}_2 \times_3 \mat{\widetilde{U}}_3 \right\rVert_{2} < \varepsilon \left\lVert \mathcal{T} \right\rVert_{2}$.}
\end{figure}

\section{Results}

In all the examples, densities and SMs obtained via Regularized Nonlinear Inversion will be labelled as "MPI-BL" (MPI BiLinear), while densities obtained via Regularized Density Reconstruction via Constrained Optimization as  "MPI-L" (MPI Linear).

\subsection{2D Cartesian sequences}
Figures 4 and 5 depict the extracted SMs and density of an MPI reconstructed axial slice from a Cartesian acquisition of a human brain obtained from the fastMRI database \cite{fastmri_web, Zbontar2018}. The data is a fully sampled Flash acquisition (TR/TE=250/3.4 ms, FA=70$^\circ$, matrix size: 320x320, slice thickness: 5mm) with a FOV of 220x220 $\text{mm}^2$, acquired at 3T using a 16-channel head coil. The data is retrospectively downsampled, according to different Cartesian undersampling patterns and ACS regions. The stability of the recovered SMs for different combination of R and ACS regions proves the effectiveness of the physics-based regularization scheme. Aliasing artifacts are visible in the final reconstructed image for acceleration factors $R \ge 4$, and substantial stability of the image is observed for ACS lines $\ge 4$. Computation time on an Intel Xeon CPU E5-2650 with NVIDIA Tesla K80 GPU is 346s and 36s for SMs and image reconstruction (R=2, ACS=16), respectively.
Figure 6 investigates the performances of MPI for simultaneous Cartesian accelerations along phase and slice directions. The dataset is a fully sampled BRAVO acquisition (TR/TE=9.972/3.92 ms, FA=10$^\circ$, matrix size: 192x192x170) with a FOV of 240x240x204 $\text{mm}^3$, acquired with a 1.5T GE using a 12-channel head coil. The frequency encoding is resolved and one single axial slice is reconstructed: different Cartesian downsampling schemes are retrospectively applied, with fixed number of ACS lines (16) and Maxwell basis with $q=50$. The reconstructed image is free from artifacts for combined $R \le 6$.

\begin{figure}[!htbp]
\centering
    {
   \includegraphics[width=\linewidth]{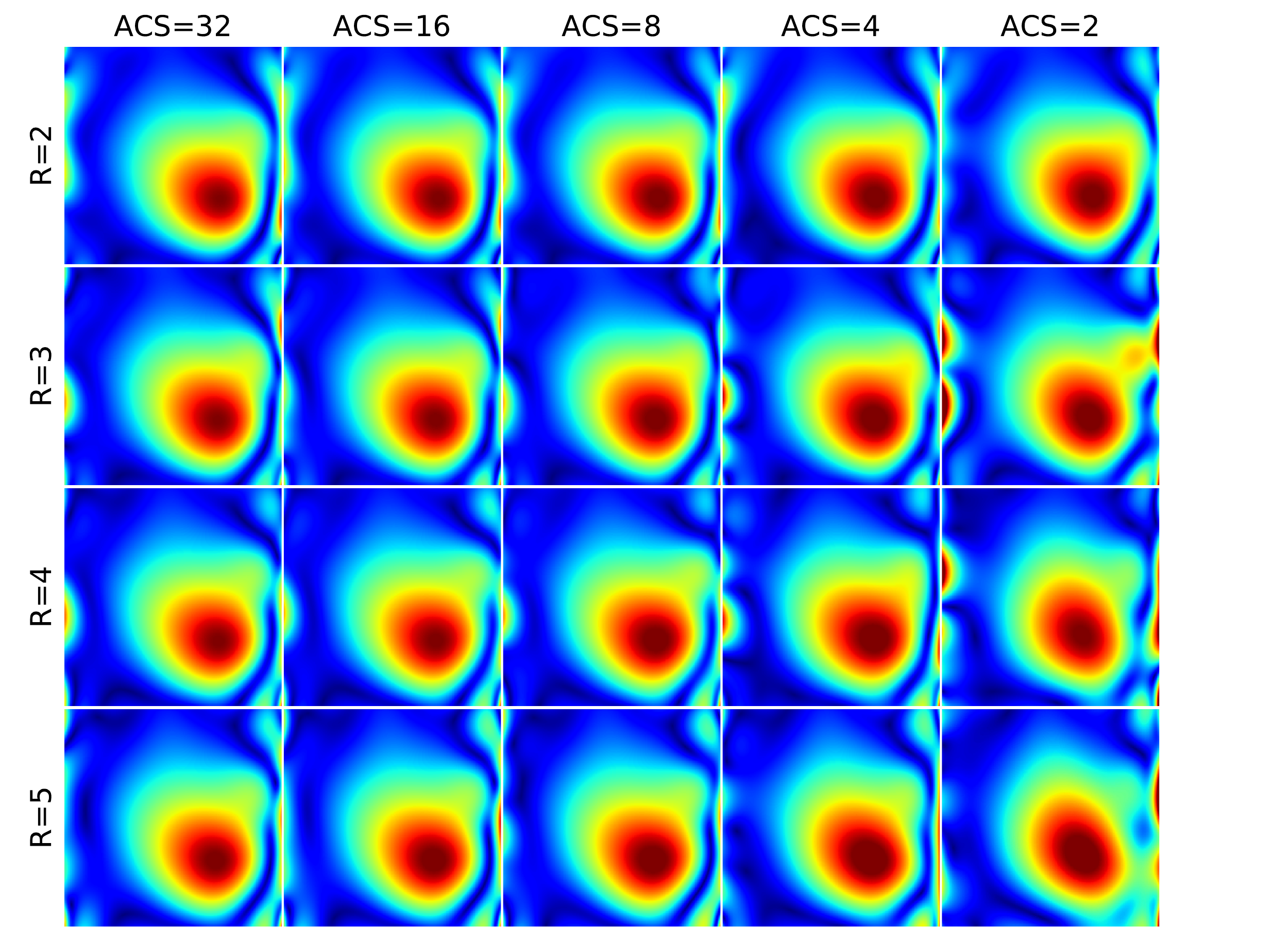}
    }
\caption{A 3T fully sampled Cartesian acquisition of a human head is retrospectively downsampled, with different Cartesian undersampling factors R and ACS regions. MPI-BL SMs extraction is carried out with a fixed basis dimension $q=100$: the magnitude of the extracted SM of one coil is shown for increasing acceleration (top to bottom) and decreasing ACSs (left to right).}
\end{figure}
%

\begin{figure}[!htbp]
\centering
    {
   \includegraphics[width=\linewidth]{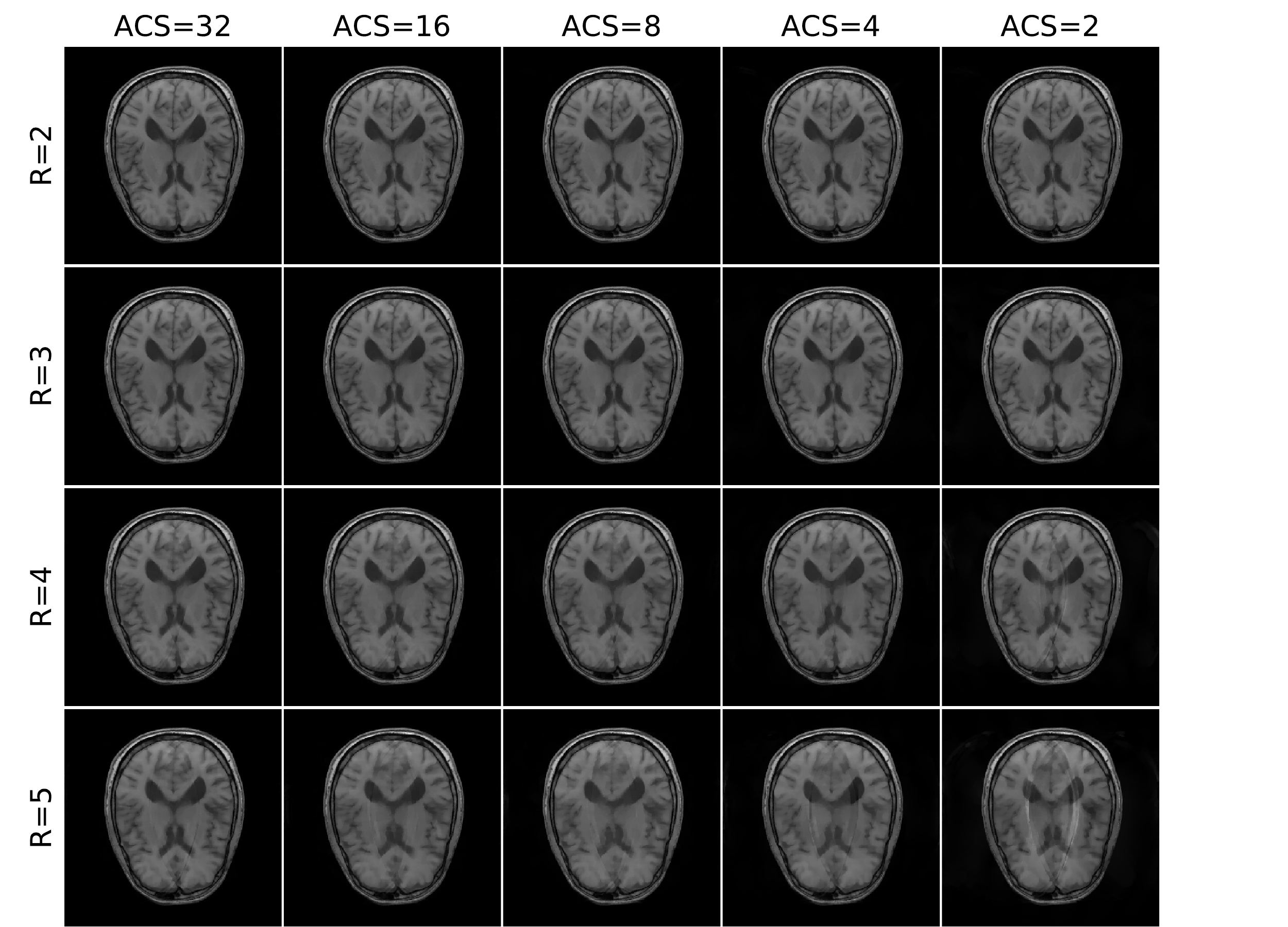}
    }
\caption{The same dataset and undersampling strategy as in Fig.~4 and the corresponding extracted SMs are used to reconstruct the image with MPI-L for increasing Cartesian downsampling R (top to bottom) and decreasing number of ACSs (left to right).}
\end{figure}
%

\begin{figure}[!htbp]
\centering
    {
   \includegraphics[width=\linewidth]{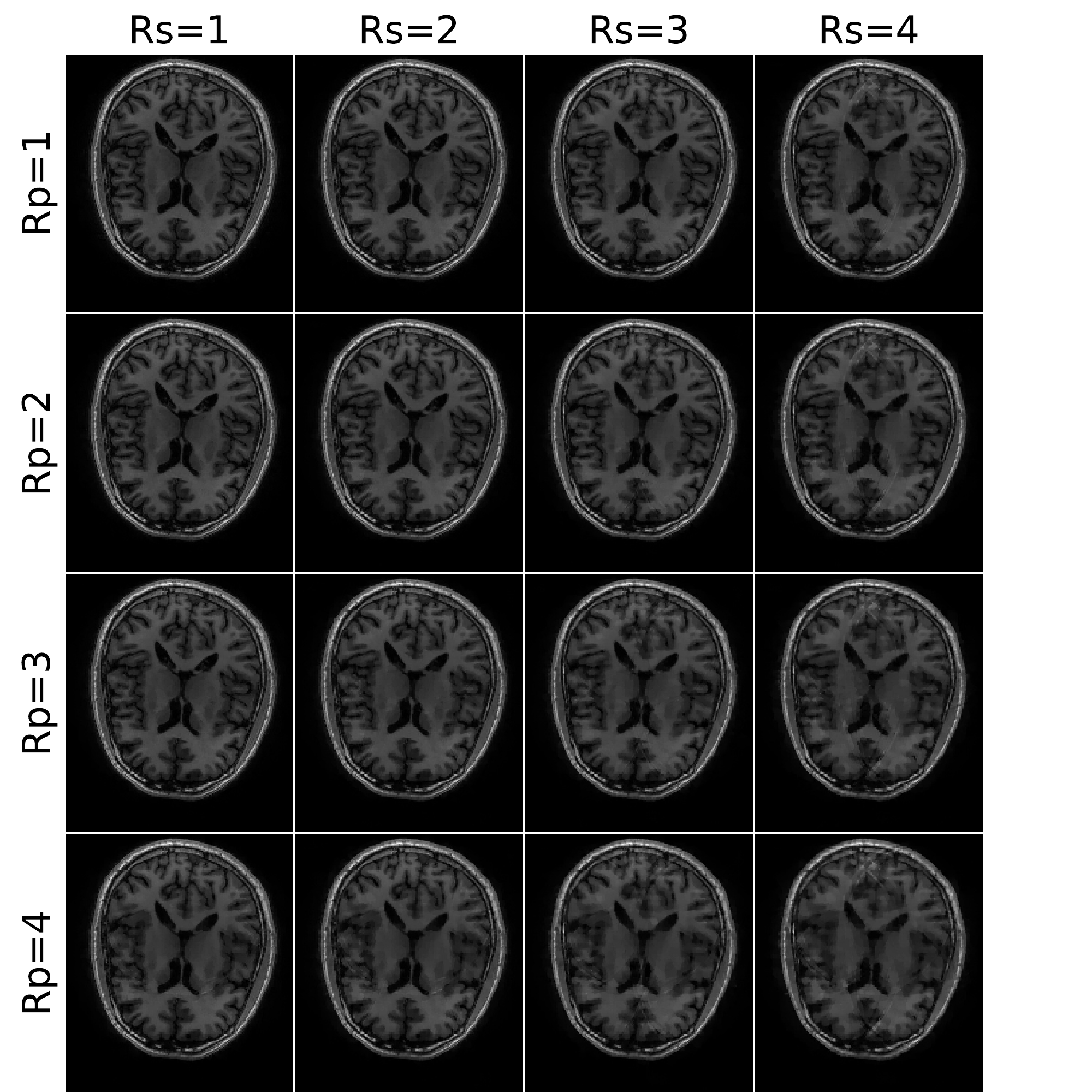}
    }
\caption{A 1.5T fully sampled Cartesian acquisition of a human head is retrospectively downsampled, with different Cartesian undersampling factors along the phase encoding (Rp) and slice encoding (Rs) dimensions, and a fixed number of 16 ACSs. The frequency encoding is resolved and one single axial slice is reconstructed as a 2D problem. The reconstructed MPI-L density is shown, with SMs extracted via MPI-BL and basis dimension $q=50$. }
\end{figure}

\subsection{2D synthetic radial sequences}
Figure 7 shows the capability of MPI to address non-Cartesian acquisitions. Provided that the operator $\mat{F}$ of Eq.~\eqref{eq:forward_model} is available, the described formulation is directly applicable. A synthetic 8 channels acquisition with golden angle radial sampling is generated from a Shepp-Logan model of size 256x256, in the presence of additive white Gaussian noise independent for each coil.

\begin{figure}[!htbp]
\centering
    {
   \includegraphics[width=\linewidth]{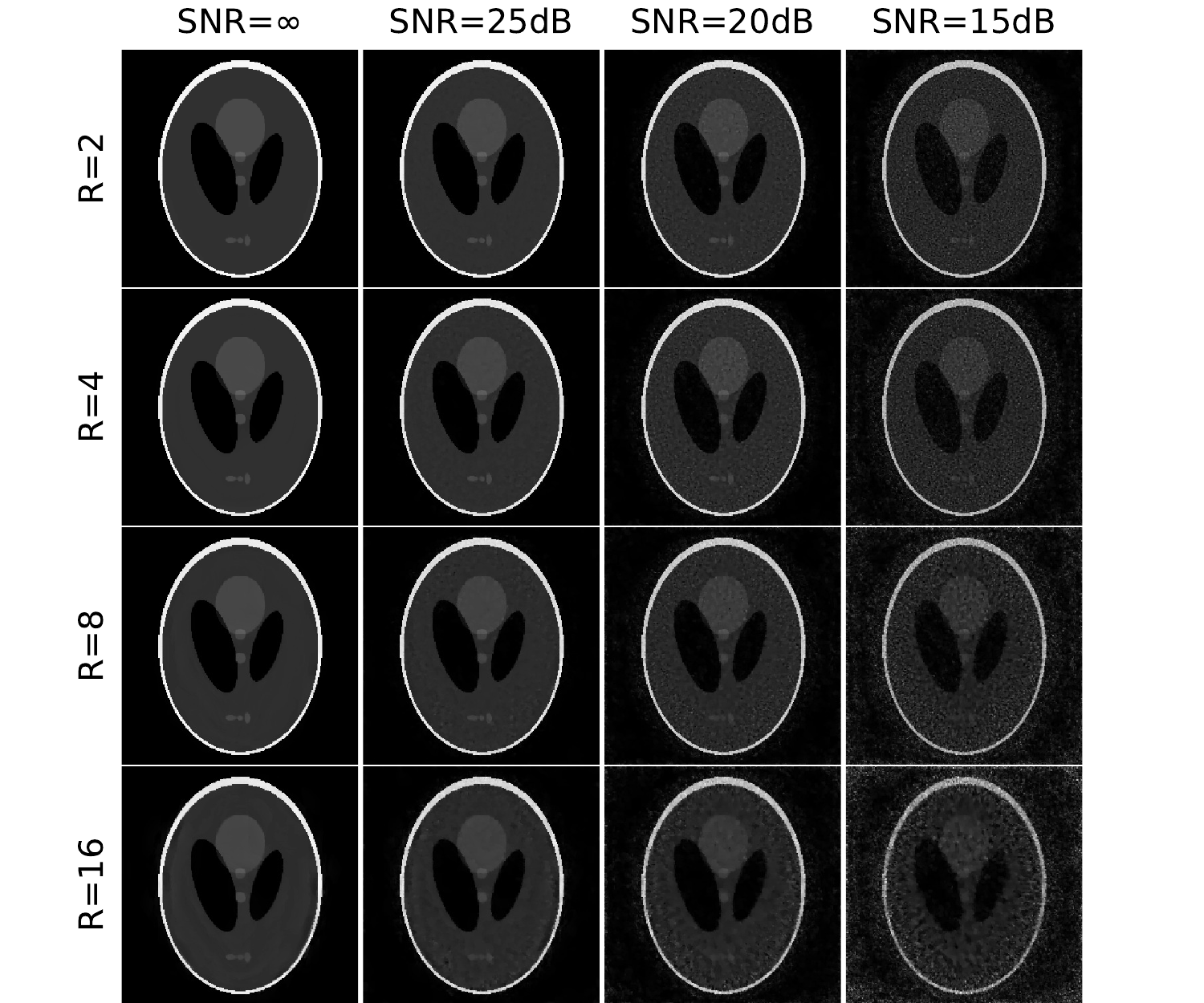}
    }
\caption{A 256x256 pixels Shepp-Logan phantom and a set of 8 synthetic SMs with the same size are used to simulate a synthetic golden angle acquisition with readout length 256. Independent white Gaussian noise is then added to the simulated $k$-space signal of each coil, yielding decreasing SNR values $\left[ \infty, 25\text{dB}, 20\text{dB}, 15\text{dB} \right]$ (left to right). The solution of MPI-L is shown for different numbers of acquired spokes $N \in \left[ 200, 100, 50, 25 \right]$, corresponding to acceleration factors $R \in \left[ 2, 4, 8, 16 \right]$ (top to bottom).
The basis dimension is set to $q=100$ for all cases.}
\end{figure}

\subsection{Comparison with previous studies}
Figure 8 explores variable-density Poisson-disc undersampled reconstructions of a knee, comparing MPI with ENLIVE and SAKE. All methods provide artifact-free reconstructions up to acceleration R=3, with the denoising step performed by MPI-L providing a generally cleaner image. For higher accelerations (R=5) SAKE misses signal from the center of the image, ENLIVE and MPI-BL both provide a rather noisy image, while the MPI-L reconstructed image has significant better quality. Figure 9 shows Cartesian reconstructions with CAIPIRINHA patterns with different acceleration factors and 24 ACS lines, with comparisons to ESPIRiT and ENLIVE. All images appear free from artifacts even at R=16. 

\begin{figure}[!htbp]
\centering
    {
   \includegraphics[width=\linewidth]{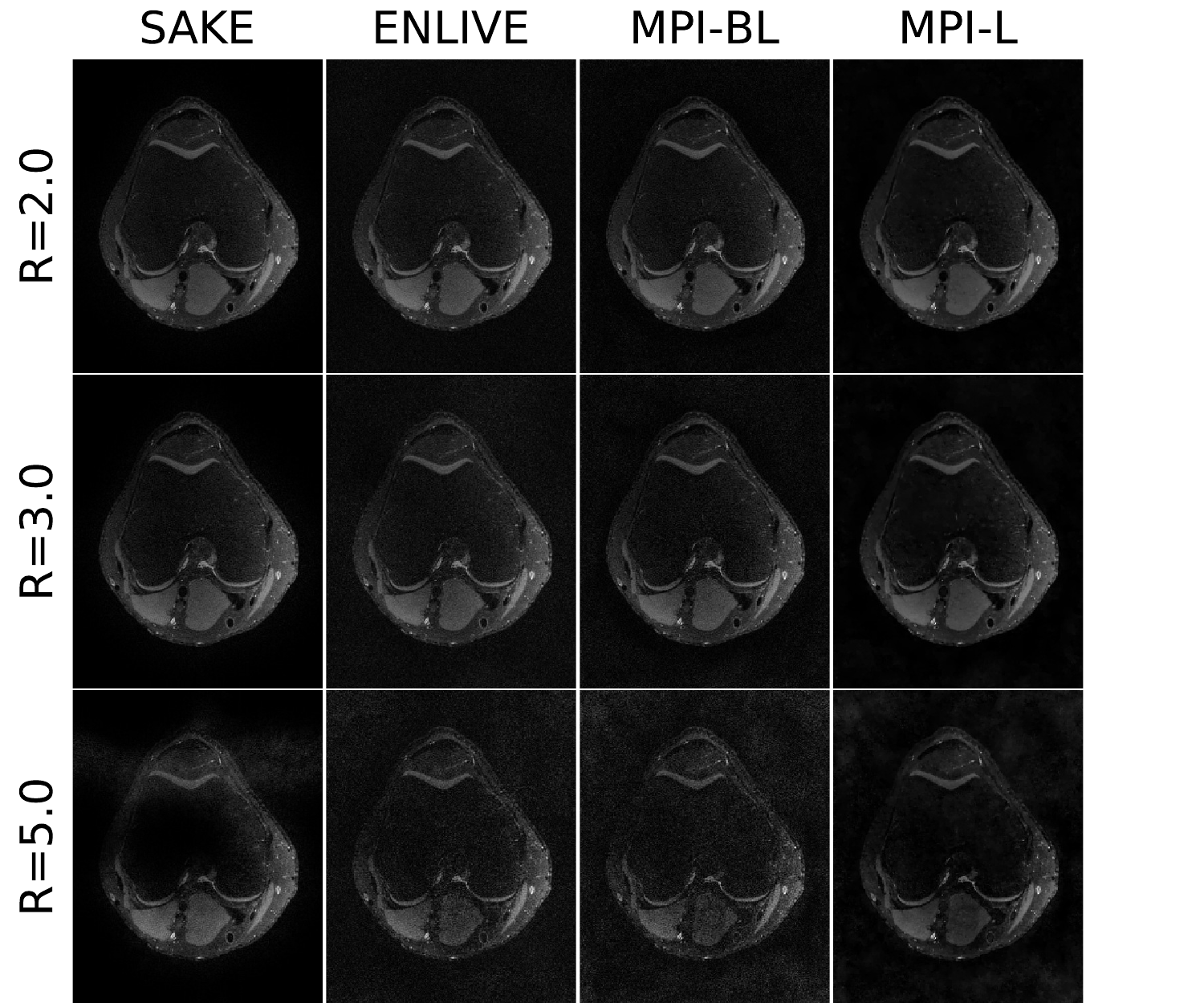}
    }
\caption{Variable-density Poisson-disc undersampled data of a human knee with varying undersampling factors reconstructed with SAKE~\cite{Shin2014}, ENLIVE~\cite{Holme2019}, MPI-BL, and MPI-L. 
All methods generate images free from artifacts up to acceleration R=3, with MPI-L providing better SNRs. For highly accelerated sequences (R=5), ENLIVE and MPI-BL provide similar images with high noise, while SAKE misses signal in the center of the image. MPI-L can provide a cleaner image even at R=5.}
\end{figure}
%

\begin{figure}[!htbp]
\centering
    {
   \includegraphics[width=\linewidth]{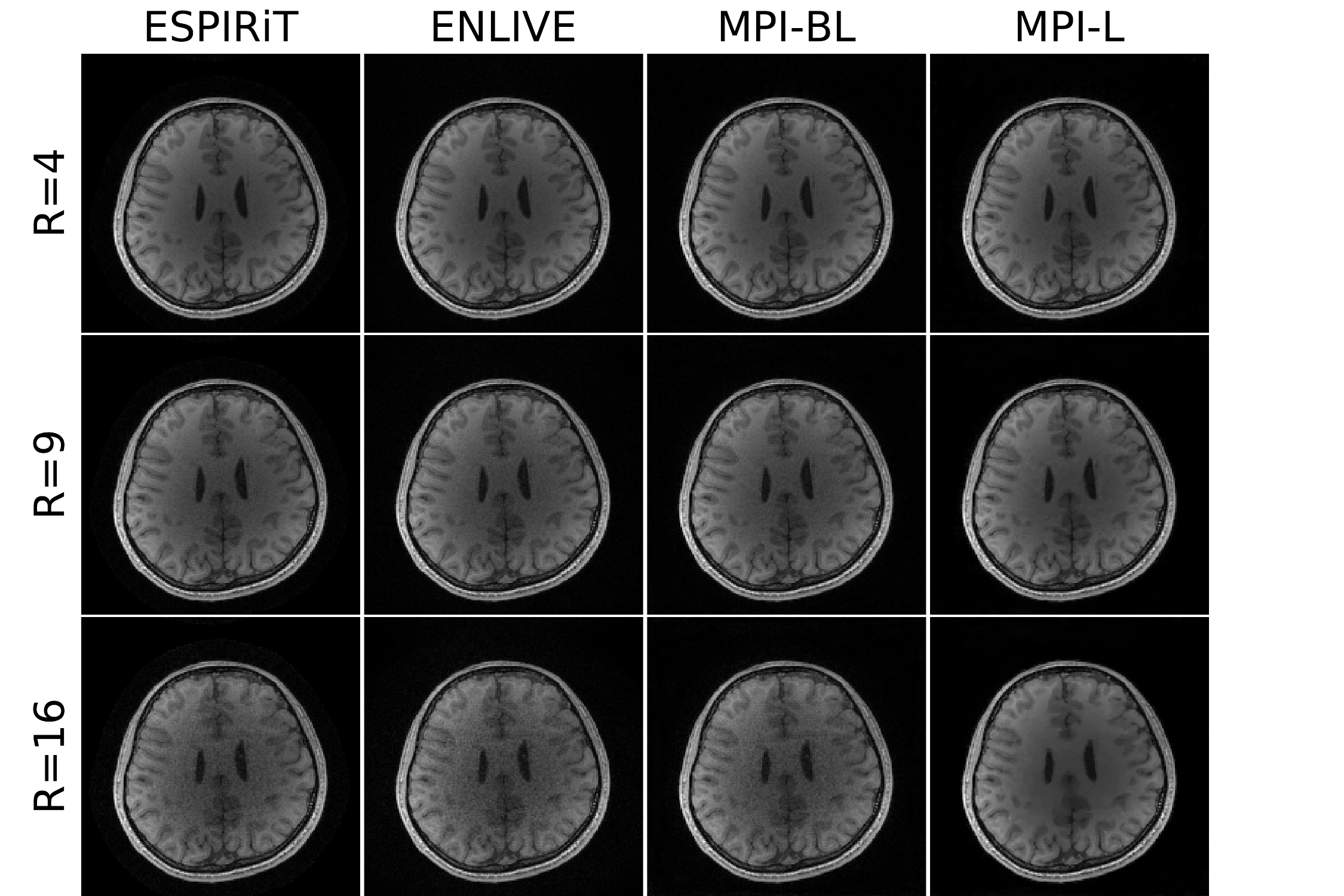}
    }
\caption{Comparison of MPI (MPI-BL and MPI-L) with ESPIRiT\cite{Uecker2014} reconstruction using 2 maps and ENLIVE~\cite{Holme2019} reconstruction using 2 maps of a single slice of a human head, undersampled with Cartesian CAIPIRINHA patterns with differing undersampling factors R and fixed size of the ACS region (24). 
The MPI-BL joint density (third column) and SMs reconstruction is obtained with a basis of dimension $q=200$; the reconstructed SMs are subsequently used as an input for the MPI-L density reconstruction (last column).}
\end{figure}

\subsection{3D sequence}
Figure 10 shows the results of a full 3D reconstruction of the same dataset of Figure 6, undersampled with a combined acceleration factor 4 and 16 ACS lines, and reconstructed with MPI-BL with a variable dimension of the Maxwell basis. The results show the flexibility of the formulation in seamlessly addressing 3D $k$-spaces with the same formulation. At $q=50$ aliasing artifacts are visible, as highlighted by the yellow arrow. When the basis is enlarged to capture these artifacts, MPI-BL yields artifact-free images. Additionally, Tucker compression reduces the memory footprint of the Maxwell basis from 19.1GB to 31MB when $q=200$, in turn enabling accelerated computations on GPU (see also Supporting Information Table S1). Computation time on an Intel Xeon E5-2686 CPU with NVIDIA Tesla V100 GPU is 73 minutes for 9 iterations of MPI-BL and $q=200$.

\begin{figure}[!htbp]
\centering
    {
   \includegraphics[width=\linewidth]{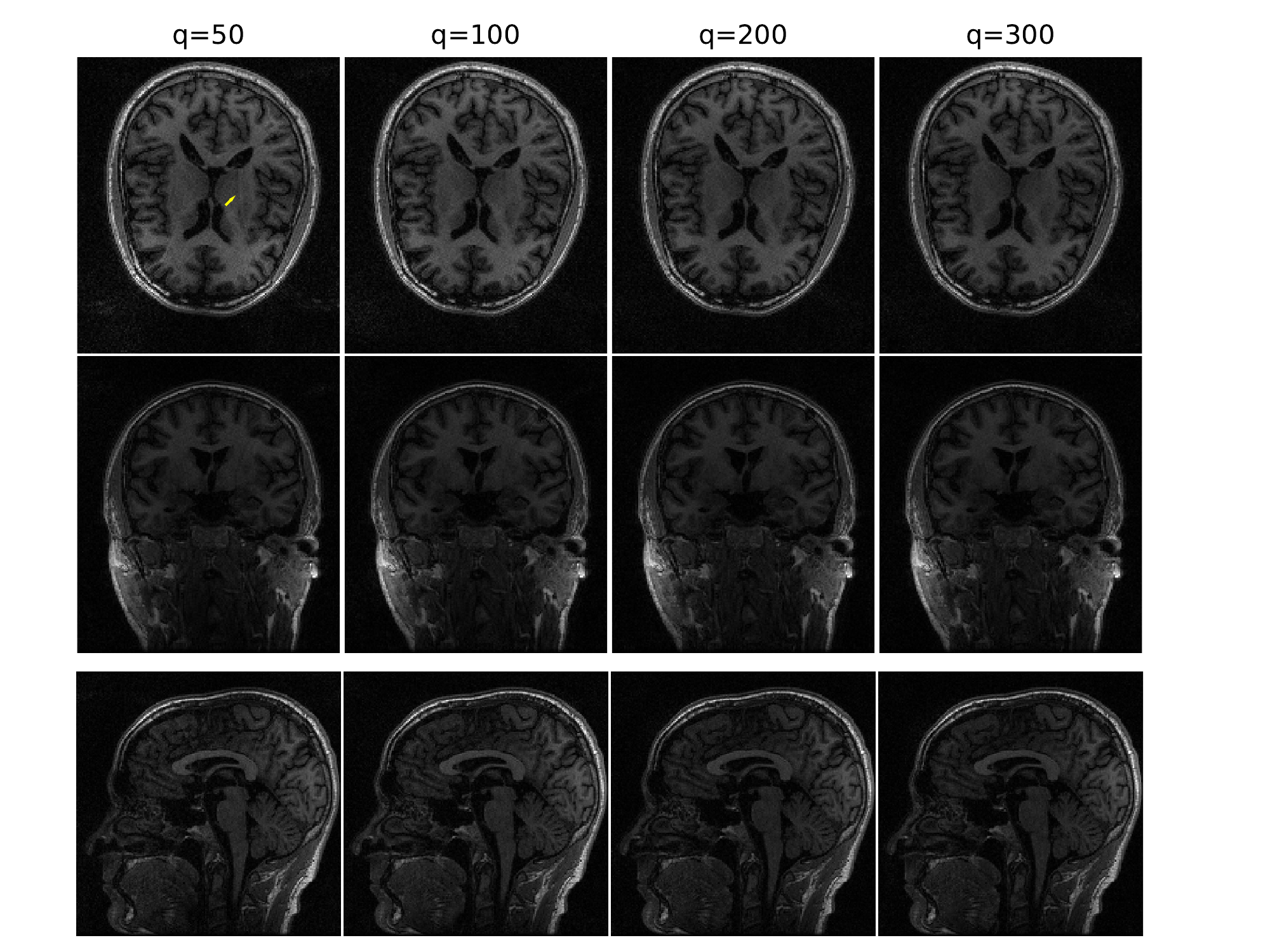}
    }
\caption{Full 3D reconstruction with MPI-BL of the same dataset of Figure 6, with cartesian undersampling factor 2 along the phase encoding and 2 along the slice encoding dimensions, corresponding to a combined acceleration 4, and 16 ACSs, for different dimensions $q$ of the basis along the columns. The full 3D density is reconstructed at once using MPI-BL with Tucker compression of the Maxwell basis: slices of the solution along the axial (top row), coronal (middle row) and sagittal (bottom row) are shown.}
\end{figure}

\section{Discussion}

\subsection{Forward Model Extensions}

There are cases where the bilinear form of the MR signal fails to capture accurately the underlying physics. More specifically, it is well documented that the image-domain methods, with the exception of ENLIVE~\cite{Holme2019}, produce erroneous results when the chosen FOV does not include entirely the object under study. As mentioned above, the SMs are essentially the circularly polarized magnetic fields received by the coils, and due to the nature of Maxwell equations their values depend strongly on the EM properties of the entire object, not only the portion inside the FOV. Hence the estimation of the actual SMs for small FOVs is an ill-posed problem. Fortunately, MPI allows the extension of the original signal equation, much like ENLIVE, with the addition of extra bilinear terms, resulting in a fairly accurate approximation of the governing physics, though in this case the estimated SMs do not correspond anymore to the true magnetic field distributions and should be considered as merely dummy variables. Nevertheless, the image reconstruction is devoid of artifacts, as evinced by the Supporting Information Figure S2, where MPI with 2 sets of maps is applied on a dataset from Ref. \cite{Holme2019}.

\subsection{Maxwell-Constrained Deep CNNs}

In recent years we have witnessed some dramatic developments in the field of machine and deep learning, where Deep Convolutional Neural Networks (DCNNs) have shown superior performance over more traditional methods in various image reconstruction tasks, such as denoising~\cite{Schmidt2014, Lefkimmiatis2018C}, demosaicking~\cite{Kokkinos2019J}, super-resolution~\cite{Haris2018}, etc. Consequently, this has also lead to an increased interest in the development of deep learning methods that could efficiently tackle the problem of MRI reconstruction~\cite{Lee2017, Knoll2020}. While in this work we have intentionally focused on an optimization-based reconstruction approach, we are convinced that a very promising future research direction, which could lead to further improvements in the reconstruction quality and offer additional robustness, is the design of physics-constrained deep reconstruction networks. The main idea here is that by constraining the solution space of a neural network, we can gain more control on the reconstruction outcome and reduce the risk of introducing erroneous reconstruction artefacts, which are completely undesirable in medical applications. In this direction, and following the discussion on the construction of the Maxwell basis, one possible way to enforce such kind of physics-based constraints is to combine the implicit Maxwell regularization approach with a variational-inspired deep network such as those introduced in~\cite{Kokkinos2019J,Kokkinos2019C}. This way, we can learn more meaningful and accurate representations for the SMs, which in turn can lead to better and more robust reconstruction results. At the same time, it is expected that by providing more information to the network about the space of solutions, we can avoid its overfitting during training and further require less training data.

\section{Conclusions}
\label{sec:conclusion}

In this work, we described a general framework for the joint reconstruction of PI data. The proposed framework introduces an expressive, physics-based regularizer for the estimation of the SMs and a constrained optimization scheme for the subsequent parameter-free density reconstruction, for improved image quality. In addition, the use of a Maxwell basis for the expansion of the SMs reduces dramatically the overall number of the unknowns in the inverse problem and accelerates the convergence of the iterative reconstruction. Finally, we utilized some relatively modern tensor decomposition methods in order to reduce the memory footprint of the Maxwell basis, which can become prohibitively large for high-resolution 3D scans. We expect this framework to allow MRI scientists and practitioners to obtain images of higher quality from datasets with even more aggressive acceleration, while its extensions in combination with deep learning-based reconstructions to offer a paradigm shift in next-generation data-driven PI approaches. 

\section*{Acknowledgements}

We thank Daniel Sodickson, Riccardo Lattanzi, and Thomas Witzel for useful discussions.


\bibliographystyle{IEEEtran}
\bibliography{references}

\begin{thebibliography}{10}
\providecommand{\url}[1]{#1}
\csname url@samestyle\endcsname
\providecommand{\newblock}{\relax}
\providecommand{\bibinfo}[2]{#2}
\providecommand{\BIBentrySTDinterwordspacing}{\spaceskip=0pt\relax}
\providecommand{\BIBentryALTinterwordstretchfactor}{4}
\providecommand{\BIBentryALTinterwordspacing}{\spaceskip=\fontdimen2\font plus
\BIBentryALTinterwordstretchfactor\fontdimen3\font minus
  \fontdimen4\font\relax}
\providecommand{\BIBforeignlanguage}[2]{{%
\expandafter\ifx\csname l@#1\endcsname\relax
\typeout{** WARNING: IEEEtran.bst: No hyphenation pattern has been}%
\typeout{** loaded for the language `#1'. Using the pattern for}%
\typeout{** the default language instead.}%
\else
\language=\csname l@#1\endcsname
\fi
#2}}
\providecommand{\BIBdecl}{\relax}
\BIBdecl

\bibitem{Sodickson1997}
D.~K. Sodickson and W.~J. Manning, ``Simultaneous acquisition of spatial
  harmonics {(SMASH)}: Fast imaging with radiofrequency coil arrays,''
  \emph{Magn. Reson. Med}, vol.~38, no.~4, pp. 591--603, 1997.

\bibitem{Pruessmann1999}
K.~P. Pruessmann, M.~Weiger, M.~B. Scheidegger, and P.~Boesiger, ``{SENSE:
  Sensitivity encoding for fast MRI},'' \emph{Magn. Reson. Med}, vol.~42,
  no.~5, pp. 952--962, 1999.

\bibitem{Griswold2002}
M.~A. Griswold, P.~M. Jakob, R.~M. Heidemann, M.~Nittka, V.~Jellus, J.~Wang,
  B.~Kiefer, and A.~Haase, ``{Generalized autocalibrating partially parallel
  acquisitions (GRAPPA)},'' \emph{Magn. Reson. Med}, vol.~47, no.~6, pp.
  1202--1210, 2002.

\bibitem{Larkman2007}
D.~J. Larkman and R.~G. Nunes, ``Parallel magnetic resonance imaging,''
  \emph{Phys. Med. Biol.}, vol.~52, no.~7, pp. 15--55, Mar 2007.

\bibitem{Ying2007}
L.~Ying and J.~Sheng, ``{Joint image reconstruction and sensitivity estimation
  in SENSE (JSENSE)},'' \emph{Magn. Reson. Med}, vol.~57, no.~6, pp.
  1196--1202, 2007.

\bibitem{Uecker2008}
M.~Uecker, T.~Hohage, K.~T. Block, and J.~Frahm, ``{Image reconstruction by
  regularized nonlinear inversion--Joint estimation of coil sensitivities and
  image content},'' \emph{Magn. Reson. Med}, vol.~60, no.~3, pp. 674--682,
  2008.

\bibitem{Holme2019}
H.~C.~M. Holme, S.~Rosenzweig, F.~Ong, R.~N. Wilke, M.~Lustig, and M.~Uecker,
  ``{ENLIVE: An efficient nonlinear method for calibrationless and robust
  parallel imaging},'' \emph{Sci Rep}, vol.~9, p. 3034, 2019.

\bibitem{Trzasko2011}
J.~D. {Trzasko} and A.~{Manduca}, ``{A Calibrationless parallel MRI using
  CLEAR},'' \emph{In Conf. Rec. Asilomar Conf. Signals Syst. Comput.}, no.~45,
  pp. 75--79, 2011.

\bibitem{Shin2014}
P.~J. Shin, P.~E.~Z. Larson, M.~A. Ohliger, M.~Elad, J.~M. Pauly, D.~B.
  Vigneron, and M.~Lustig, ``{Calibrationless parallel imaging reconstruction
  based on structured low-rank matrix completion},'' \emph{Magn. Reson. Med},
  vol.~72, no.~4, pp. 959--970, 2014.

\bibitem{Haldar2014}
J.~P. Haldar, ``{Low-rank modeling of local $k$-space neighborhoods (LORAKS)
  for constrained MRI},'' \emph{IEEE Trans. Med. Imag.}, vol.~33, no.~3, pp.
  668--681, Mar. 2014.

\bibitem{Haldar2016}
J.~P. Haldar and J.~Zhuo, ``{P-LORAKS: Low-rank modeling of local $k$-space
  neighborhoods with parallel imaging data},'' \emph{Magn. Reson. Med},
  vol.~75, no.~4, pp. 1499--1514, 2016.

\bibitem{Lustig2007}
M.~Lustig, D.~Donoho, and J.~M. Pauly, ``{Sparse MRI: The application of
  compressed sensing for rapid MR imaging},'' \emph{Magn. Reson. Med}, vol.~58,
  no.~6, pp. 1182--1195, 2007.

\bibitem{Huang2010}
F.~Huang, Y.~Chen, W.~Yin, W.~Lin, X.~Ye, W.~Guo, and A.~Reykowski, ``{A rapid
  and robust numerical algorithm for sensitivity encoding with sparsity
  constraints: Self-feeding sparse SENSE},'' \emph{Magn. Reson. Med}, vol.~64,
  no.~4, pp. 1078--1088, 2010.

\bibitem{Knoll2012}
F.~Knoll, C.~Clason, K.~Bredies, M.~Uecker, and R.~Stollberger, ``Parallel
  imaging with nonlinear reconstruction using variational penalties,''
  \emph{Magn. Reson. Med}, vol.~67, no.~1, pp. 34--41, 2012.

\bibitem{Knoll2020}
F.~Knoll, K.~Hammernik, C.~Zhang, S.~Moeller, T.~Pock, D.~K. Sodickson, and
  M.~Akcakaya, ``Deep-learning methods for parallel magnetic resonance imaging
  reconstruction: A survey of the current approaches, trends, and issues,''
  \emph{IEEE Signal Processing Magazine}, vol.~37, no.~1, pp. 128--140, 2020.

\bibitem{Morrison2007}
R.~L. {Morrison}, M.~{Jacob}, and M.~N. {Do}, ``{Multichannel estimation of
  coil sensitivities in parallel MRI},'' \emph{{In Proceedings of the 4th IEEE
  International Symposium on Biomedical Imaging: From Nano to Macro}}, pp.
  117--120, April 2007.

\bibitem{Jin2010}
J.~Jin, F.~Liu, E.~Weber, Y.~Li, and S.~Crozier, ``{An electromagnetic reverse
  method of coil sensitivity mapping for parallel MRI -- Theoretical
  framework},'' \emph{Journal of Magnetic Resonance}, vol. 207, no.~1, pp. 59
  -- 68, 2010.

\bibitem{Allison2012}
M.~J. {Allison}, S.~{Ramani}, and J.~A. {Fessler}, ``{Regularized MR coil
  sensitivity estimation using augmented Lagrangian methods},'' \emph{In
  Proceedings of the 9th IEEE International Symposium on Biomedical Imaging,
  Barcelona, Spain}, pp. 394--397, May 2012.

\bibitem{Uecker2014}
M.~Uecker, P.~Lai, M.~J. Murphy, P.~Virtue, M.~Elad, J.~M. Pauly, S.~S.
  Vasanawala, and M.~Lustig, ``{ESPIRiT--an eigenvalue approach to
  autocalibrating parallel MRI: Where SENSE meets GRAPPA},'' \emph{Magn. Reson.
  Med}, vol.~71, no.~3, pp. 990--1001, 2014.

\bibitem{Ma2015}
Y.-J. Ma, W.~Liu, X.~Tang, and J.-H. Gao, ``Improved {SENSE} imaging using
  accurate coil sensitivity maps generated by a global magnitude-phase fitting
  method,'' \emph{Magn. Reson. Med}, vol.~74, no.~1, pp. 217--224, 2015.

\bibitem{Zbontar2018}
J.~Zbontar, F.~Knoll, A.~Sriram, and et~al., ``{fastMRI: An open dataset and
  benchmarks for accelerated MRI},'' \emph{arXiv:1811.08839 preprint.}, 2018.

\bibitem{Hansen2006}
P.~C. Hansen, J.~G. Nagy, and D.~P. O'Leary, \emph{Deblurring Images: Matrices,
  Spectra, and Filtering}.\hskip 1em plus 0.5em minus 0.4em\relax SIAM, 2006.

\bibitem{Bakushinsky2005}
A.~B. Bakushinsky and M.~Y. Kokurin, \emph{Iterative methods for approximate
  solution of inverse problems}.\hskip 1em plus 0.5em minus 0.4em\relax
  Springer, 2005, vol. 577.

\bibitem{Levin2011}
A.~Levin, Y.~Weiss, F.~Durand, and W.~T. Freeman, ``Understanding blind
  deconvolution algorithms,'' \emph{IEEE transactions on pattern analysis and
  machine intelligence}, vol.~33, no.~12, pp. 2354--2367, 2011.

\bibitem{Shewchuk1994}
\BIBentryALTinterwordspacing
J.~R. Shewchuk, ``An introduction to the conjugate gradient method without the
  agonizing pain,'' 1994. [Online]. Available:
  \url{http://www.cs.cmu.edu/~jrs/jrspapers.html}
\BIBentrySTDinterwordspacing

\bibitem{Rudin1992}
L.~Rudin, S.~Osher, and E.~Fatemi, ``Nonlinear total variation based noise
  removal algorithms,'' \emph{Physica D}, vol.~60, pp. 259--268, 1992.

\bibitem{Lefkimmiatis2015Ja}
S.~Lefkimmiatis, A.~Roussos, P.~Maragos, and M.~Unser, ``Structure tensor total
  variation,'' \emph{SIAM Journal on Imaging Sciences}, vol.~8, no.~2, pp.
  1090--1122, 2015.

\bibitem{Lefkimmiatis2015Jb}
S.~Lefkimmiatis and S.~Osher, ``Nonlocal structure tensor functionals for image
  regularization,'' \emph{IEEE Transactions on Computational Imaging}, vol.~1,
  no.~1, pp. 16--29, March 2015.

\bibitem{Lefkimmiatis2013J}
S.~Lefkimmiatis, J.~Ward, and M.~Unser, ``{Hessian} {Schatten}-norm
  regularization for linear inverse problems,'' \emph{IEEE Trans. Image
  Process.}, vol.~22, no.~5, pp. 1873--1888, 2013.

\bibitem{Esser2009}
E.~Esser, ``Applications of {Lagrangian}-based alternating direction methods
  and connections to split {Bregman},'' \emph{CAM report}, vol.~9, 2009.

\bibitem{Boyd2011}
S.~Boyd, N.~Parikh, E.~Chu, B.~Peleato, and J.~Eckstein, \emph{Distributed
  Optimization and Statistical Learning via the Alternating Direction Method of
  Multipliers}.\hskip 1em plus 0.5em minus 0.4em\relax Now Publishers, 2011.

\bibitem{Lefkimmiatis2013Jb}
S.~Lefkimmiatis and M.~Unser, ``{Poisson} image reconstruction with {Hessian}
  {Schatten}-norm regularization,'' \emph{IEEE Trans. Image Process.}, vol.~22,
  pp. 4314--4327, 2013.

\bibitem{Combettes2005}
P.~L. Combettes and V.~R. Wajs, ``Signal recovery by proximal forward-backward
  splitting,'' \emph{Multiscale Model. Simul.}, vol.~4, no.~4, pp. 1168--1200,
  2005.

\bibitem{He2000}
B.~He, H.~Yang, and S.~Wang, ``Alternating direction method with self-adaptive
  penalty parameters for monotone variational inequalities,'' \emph{Journal of
  Optimization Theory and applications}, vol. 106, no.~2, pp. 337--356, 2000.

\bibitem{Blomgren1998}
P.~Blomgren and T.~F. Chan, ``Color tv: total variation methods for restoration
  of vector-valued images,'' \emph{IEEE transactions on image processing},
  vol.~7, no.~3, pp. 304--309, 1998.

\bibitem{Love1901}
A.~E.~H. Love, ``I. the integration of the equations of propagation of electric
  waves,'' \emph{Philosophical Transactions of the Royal Society of London.
  Series A, Containing Papers of a Mathematical or Physical Character}, vol.
  197, no. 287--299, pp. 1--45, 1901.

\bibitem{Ishimaru1978}
A.~Ishimaru, \emph{Wave propagation and scattering in random media}.\hskip 1em
  plus 0.5em minus 0.4em\relax Academic Press, 1978.

\bibitem{Vaidya2016}
M.~V. Vaidya, C.~M. Collins, D.~K. Sodickson, R.~Brown, G.~C. Wiggins, and
  R.~Lattanzi, ``Dependence of b1+ and b1- field patterns of surface coils on
  the electrical properties of the sample and the mr operating frequency.
  concepts in magnetic resonance.'' \emph{Concepts Magn Reson Part B Magn Reson
  Eng}, vol.~46, no.~1, pp. 25--40, 2016.

\bibitem{MARIE}
\BIBentryALTinterwordspacing
A.~G. {Polimeridis} and J.~F. {Villena}, ``{MARIE: MAgnetic Resonance Integral
  Equation suite}.'' [Online]. Available:
  \url{https://github.com/thanospol/MARIE}
\BIBentrySTDinterwordspacing

\bibitem{Polimeridis2014}
A.~Polimeridis, J.~Villena, L.~Daniel, and J.~White, ``{Stable FFT-JVIE solvers
  for fast analysis of highly inhomogeneous dielectric objects},'' \emph{J.
  Comput. Phys.}, vol. 269, pp. 280 -- 296, 2014.

\bibitem{Villena2016}
J.~F. {Villena}, A.~G. {Polimeridis}, Y.~{Eryaman}, E.~{Adalsteinsson}, L.~L.
  {Wald}, J.~K. {White}, and L.~{Daniel}, ``{Fast electromagnetic analysis of
  MRI transmit RF coils based on accelerated integral equation methods},''
  \emph{IEEE Trans. Biomed. Eng.}, vol.~63, no.~11, pp. 2250--2261, Nov. 2016.

\bibitem{Liberty20167}
E.~Liberty, F.~Woolfe, P.-G. Martinsson, V.~Rokhlin, and M.~Tygert,
  ``Randomized algorithms for the low-rank approximation of matrices,''
  \emph{Proceedings of the National Academy of Sciences}, vol. 104, no.~51, pp.
  20\,167--20\,172, 2007.

\bibitem{Halko2011}
N.~Halko, P.~G. Martinsson, and J.~A. Tropp, ``{Finding structure with
  randomness: probabilistic algorithms for constructing approximate matrix
  decompositions},'' \emph{SIAM Rev.}, vol.~53, no.~2, pp. 217--288, 2011.

\bibitem{sigpy}
\BIBentryALTinterwordspacing
Sigpy. [Online]. Available:
  \url{https://sigpy.readthedocs.io/en/latest/mri.html}
\BIBentrySTDinterwordspacing

\bibitem{Tucker1966}
L.~R. Tucker, ``Some mathematical notes on three-mode factor analysis,''
  \emph{Psychometrika}, vol.~31, pp. 279--311, 1966.

\bibitem{Giannakopoulos2019}
I.~I. {Giannakopoulos}, M.~S. {Litsarev}, and A.~G. {Polimeridis}, ``Memory
  footprint reduction for the {FFT}-based volume integral equation method via
  tensor decompositions,'' \emph{IEEE Trans. Antennas Propag.}, vol.~67,
  no.~12, pp. 7476--7486, 2019.

\bibitem{rabanser2017introduction}
S.~Rabanser, O.~Shchur, and S.~Günnemann, ``Introduction to tensor
  decompositions and their applications in machine learning,'' 2017.

\bibitem{fastmri_web}
\BIBentryALTinterwordspacing
fastmri. [Online]. Available: \url{https://fastmri.med.nyu.edu/}
\BIBentrySTDinterwordspacing

\bibitem{Schmidt2014}
U.~Schmidt and S.~Roth, ``Shrinkage fields for effective image restoration,''
  \emph{In Proc. IEEE Int. Conf. Computer Vision and Pattern Recognition
  (CVPR)}, pp. 2774--2781, 2014.

\bibitem{Lefkimmiatis2018C}
S.~Lefkimmiatis, ``Universal denoising networks : A novel cnn architecture for
  image denoising,'' \emph{In Proc. IEEE Int. Conf. Computer Vision and Pattern
  Recognition (CVPR)}, June 2018.

\bibitem{Kokkinos2019J}
F.~Kokkinos and S.~Lefkimmiatis, ``Iterative joint image demosaicking and
  denoising using a residual denoising network,'' \emph{IEEE Transactions on
  Image Processing}, vol.~28, no.~8, pp. 4177--4188, Aug 2019.

\bibitem{Haris2018}
M.~Haris, G.~Shakhnarovich, and N.~Ukita, ``Deep back-projection networks for
  super-resolution,'' \emph{In Proc. IEEE Int. Conference Computer Vision and
  Pattern Recognition (CVPR)}, pp. 1664--1673, 2018.

\bibitem{Lee2017}
D.~Lee, J.~Yoo, and J.~C. Ye, ``Deep artifact learning for compressed sensing
  and parallel mri,'' \emph{arXiv preprint arXiv:1703.01120}, 2017.

\bibitem{Kokkinos2019C}
F.~Kokkinos and S.~Lefkimmiatis, ``Iterative residual cnns for burst
  photography applications,'' \emph{In Proc. IEEE Int. Conf. Computer Vision
  and Pattern Recognition (CVPR)}, June 2019.

\end{thebibliography}

\clearpage

\section*{Supporting material}

\begin{table*}[!!t]
\centering
\begin{tabular}{||c | c c c | c c c ||} 
 \hline
  &  & Dense &  &  & Tucker &  \\ [1ex]
  \hline
 $q$ & RAM & CPU MVP & GPU MVP & RAM & CPU MVP & GPU MVP \\
  & [MB] & [sec] & [sec](*) & [MB] & [sec] & [sec](*)  \\
 \hline\hline
 75 & 7,172 & 0.18 & 0.1 & 11 & 18.9 & 0.6 \\
 200 & 19,125 & 0.35 & -- & 31 & 50.4 & 1.1 \\
 500 & 47,812 & 0.8 & -- & 129 & 127.1 & 2.9 \\
 \hline
\end{tabular}
\caption{Time and memory requirements for the basis stored as a dense matrix and in compressed form with accuracy $\epsilon_t=10^{-4}$. 
The FOV has size 192x192x170 voxels, with corresponding matrix size $N \! \times \! q , \, N=6266880$.\\
MVP: Matrix Vector Multiplication time \\
(*): computed on CUDA-enabled pyTorch code (CUDA Version: 10.1, GPU NVIDIA Tesla V100).\\
--: the data does not fit on GPU}
\end{table*}

\begin{figure}[!h]
\centering
    {
   \includegraphics[width=\linewidth]{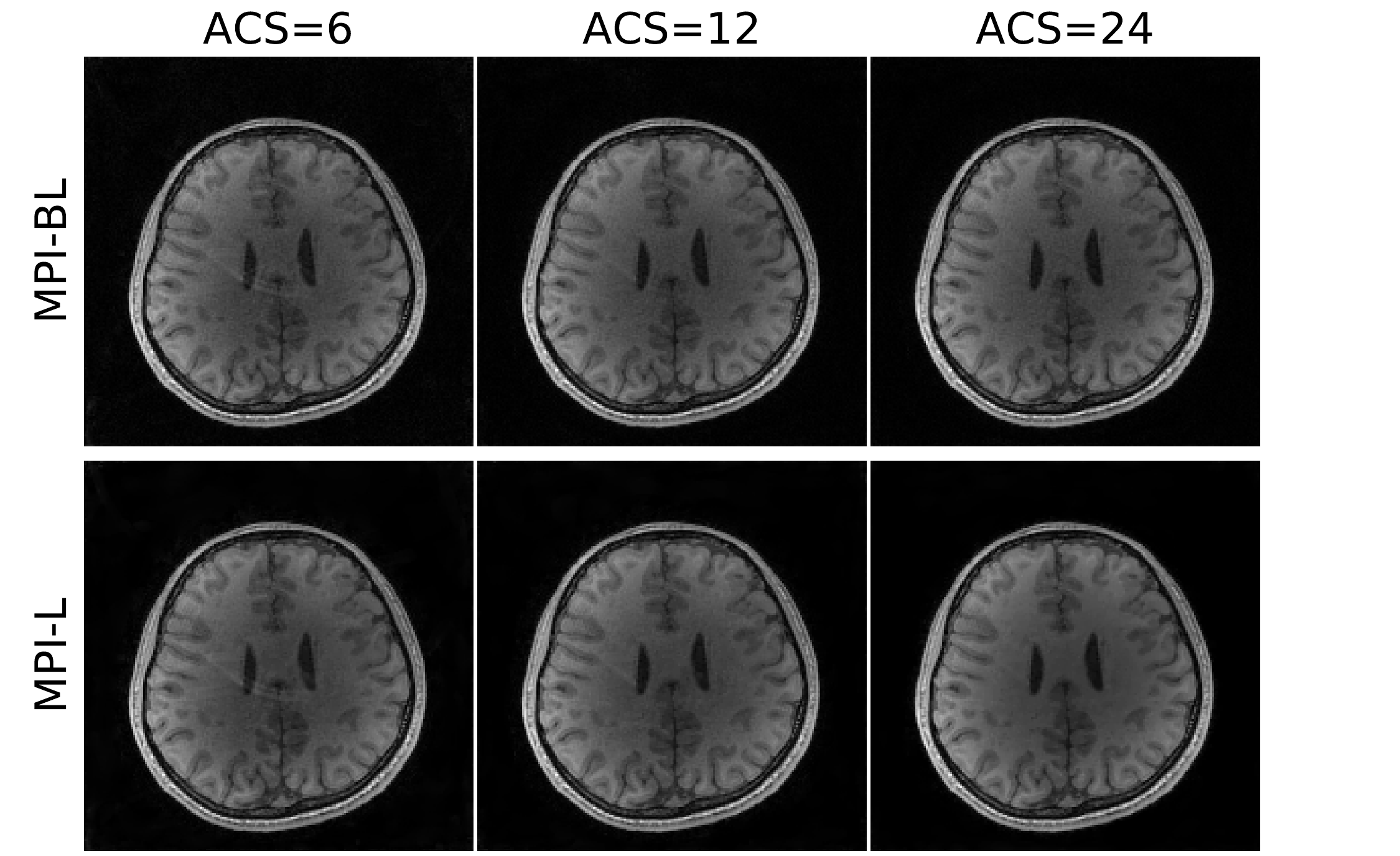}
    }
\caption{MPI reconstruction of the same dataset of Figure 9, undersampled with Cartesian CAIPIRINHA pattern with undersampling factor R=9 and and variable number of ACS lines. The results of MPI-BL (with basis dimension $q=200$) and with MPI-L demonstrate the flexibility of MPI, which does not explicitly depend on the sampling pattern, allowing to reduce the size of the low-frequency portion of $k$-space without abruptly breaking up.}
\end{figure}
%

\begin{figure}[!h]
\centering
    {
   \includegraphics[width=\linewidth]{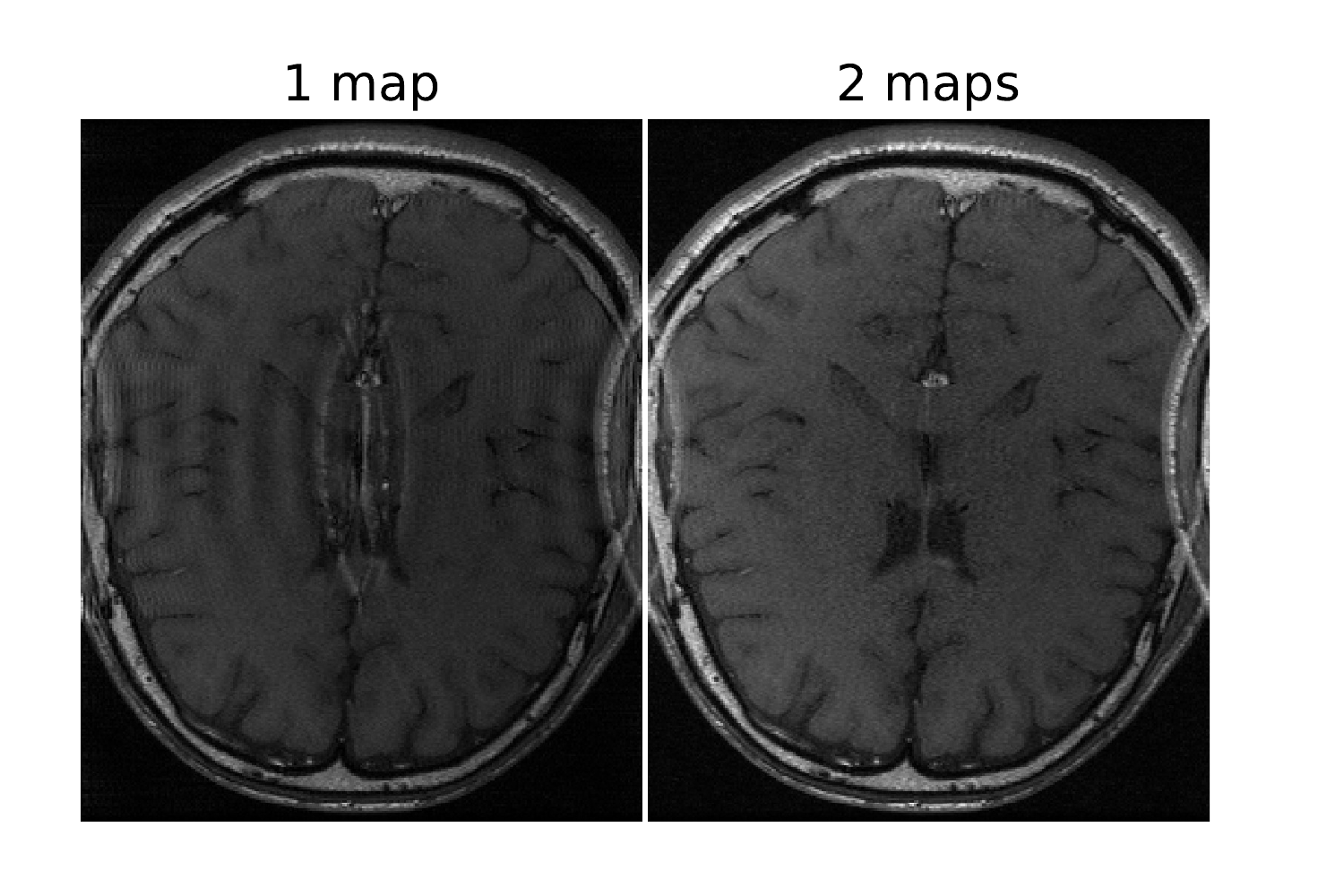}
    }
\caption{MPI reconstruction of the dataset from \cite{Uecker2014}, corresponding to a truncated FOV acquisition of a factor 2 undersampled 2D spin-echo dataset acquired at 1.5T. 19 iterations of the MPI-BL solver with basis dimension $q=50$ are used to generate solutions with 1 and 2 maps: the solution allowing 2 maps (right) is free from artifacts, which are clearly visible in the center of the single map solution (left).}
\end{figure}

\end{document}